\documentclass[twocolumn,
nofootinbib,
 amsmath,amssymb,
 aps,
 prd,
 floatfix,
 superscriptaddress
]{revtex4}
\usepackage{tabularray}
\usepackage{graphicx}% Include figure files
\usepackage{dcolumn}% Align table columns on decimal point
\usepackage{bm}% bold math
\usepackage{dsfont}
\usepackage[normalem]{ulem} 
\usepackage[center]{caption}
\usepackage{xcolor}

\usepackage{hyperref}
\usepackage{cleveref}
\usepackage{gensymb}

\begin{document}

%\preprint{APS/123-QED}

\title{Probing the axion-photon coupling with space-based gravitational wave detectors}% Force line breaks with \\

\author{Jordan Gu\'e}\email{jgue@ifae.es}
%\email{jordan.gue@obspm.fr}
\affiliation{Institut de F\'isica d’Altes Energies (IFAE), The Barcelona Institute of Science and Technology, Campus UAB, 08193 Bellaterra (Barcelona), Spain}
\affiliation{%
 SYRTE, Observatoire de Paris, Universit\'e PSL, CNRS, Sorbonne Universit\'e, LNE, 61 avenue de l’Observatoire 75014 Paris, France\\
}%
\author{Aur\'elien Hees}
\affiliation{%
 SYRTE, Observatoire de Paris, Universit\'e PSL, CNRS, Sorbonne Universit\'e, LNE, 61 avenue de l’Observatoire 75014 Paris, France\\
}%
\author{Peter Wolf}
\affiliation{%
 SYRTE, Observatoire de Paris, Universit\'e PSL, CNRS, Sorbonne Universit\'e, LNE, 61 avenue de l’Observatoire 75014 Paris, France\\
}%

\begin{abstract}
We propose a simple modification of space-based gravitational wave (GW) detector optical benches which would enable the measurement of vacuum birefringence of light induced by axion dark matter through its coupling to electromagnetism. Specifically, we propose to change a half-wave plate by a circular polarizer. While marginally affecting the sensitivity to GW by a factor $\sqrt{2}$, we show that such an adjustment would make future detectors such as LISA, TianQin, Taiji and Big-Bang Observer the most sensitive experiments at low axion masses.
\end{abstract}

\maketitle

\section{\label{sec:Intro}Introduction}

Almost $100$ years after the first hints for dark matter (DM) through its gravitational effect on visible matter \cite{Zwicky33}, its microscopic nature remains unknown and constitutes one of the biggest mysteries in modern physics \cite{Bertone18}. Several cosmological and astrophysical observations (see e.g. \cite{Battaglieri17,Carr16}) inform us on the bounds of the mass of the DM constituent, ranging from $10^{-22}$ eV to $10^{68}$ eV. These $\sim 90$ orders of magnitude of mass imply that the search for the nature of DM constitutes an experimental challenge. Out of all the DM candidates, ultralight dark matter (ULDM), with masses below $1$ eV has recently gained a lot of attention because of lack of evidence of historically dominant models such as weakly interacting massive particles. In addition, the recent extraordinary development of low-energy precision measurement techniques \cite{Safronova18} provides a hope of near-future detection for such low-mass candidates. 
One of the best-motivated models of ULDM is the axionlike-particle (in the following referred to as the axion), a pseudo-scalar DM candidate \cite{Weinberg78,Wilczek78}, which in some circumstances, can also solve the strong CP problem \cite{Peccei_Quinn77}. The coupling between such axion and electromagnetism (EM) arises naturally in many theories (see e.g. \cite{Sikivie21, Arias,Svrcek06}) and produces many different observable effects in many different systems (see \cite{Sikivie21} for a review). One of them is the phase velocity difference between the two circularly polarized states of light, known as vacuum birefringence \cite{Obata18}. In such a case, one could observe a phase shift between two circularly polarized beams with different polarizations or between a circularly polarized beam and a linearly polarized beam, whose dispersion relation is unchanged at first order. 

Searches of the axion-photon coupling through vacuum birefringence have recently been performed using an optical cavity \cite{Michimura20}. In addition, a modification to ground-based gravitational wave (GW) detectors was proposed in order to be sensitive to such an effect \cite{Nagano19}.
In this paper, we propose a new scheme to search for the axion, based on the interferometric optical system used for future space-based GW detectors, such as LISA \cite{Colpi24}, TianQin \cite{Luo16}, Taiji \cite{Luo20} and Big-Bang Observer (BBO) \cite{Crowder05}. The principle of such space-missions relies on the measurement of small variations of the optical pathlength between free falling test-masses contained in distant spacecrafts (S/C). Currently, the standard design of optical benches (OB) of those experiments makes use only of linearly polarized light. With such a setup, these space-based GW detectors are insensitive to vacuum birefringence impacting circularly polarized light such as the one induced by axions. In this paper, we propose a slight modification to the design of such OB  which would change the polarization of the laser beams exchanged between the spacecraft from linear to circular. With such a setup, the space-based GW observatory becomes sensitive to the axion-photon coupling, without jeopardizing the primary goal of the experiment, i.e detecting GW. More particularly, the proposed modification offers sensitivity to both GW and DM, whilst still satisfying the requirement that stray light from spurious reflections of the emission be rejected due to its polarization.
Assuming this minor setup modification, we estimate the sensitivity of those various experiments to the axion-photon coupling, and we show that they would all reach unprecedented sensitivities, compared to existing bounds from astrophysics and cosmology, in a mass region where laboratory experiments are poorly represented. 
Together with \cite{Nagano19}, this study completes the potential search of the axion-photon coupling through future GW detectors, and we show that ground-based and space-based detectors would be complementary for this matter.

The paper is organised as follows. In Sec.~\ref{sec:vacuum_birefringence}, we quickly derive the phase velocity difference between left and right circular polarized light induced by the axion-photon coupling. In Sec.~\ref{sec:OB}, we discuss the various requirements for GW and axion detection in terms of light polarization inside the S/C OB, and we show that a slight modification of the OB current design would make both detections possible. In Sec.~\ref{sec:results}, we compute the single-arm Doppler shift induced by the oscillation of the phase velocity of circularly polarized light, and we derive the sensitivity curves of each GW space-based detector, by making use of the appropriate Time Delay Interferometry (TDI) combination \cite{Armstrong99, Tinto99, Dhurandhar02}. Finally, we provide a short discussion of our results. 

\section{Vacuum birefringence}\label{sec:vacuum_birefringence}

In this section, we derive the vacuum birefringence arising from the axion-photon coupling with strength $g_{a\gamma}$ (in units of GeV$^{-1}$). 
We start from the action of the dimensionless axion pseudo-scalar field $a$ with mass $m_a$ that we define as
\begin{align}
    S_a &= \frac{1}{c}\int d^4x \sqrt{-g}\left[\frac{R}{2\kappa} - \frac{1}{2\kappa}g^{\mu\nu}\partial_\mu a \partial_\nu a -\frac{m^2_a c^2}{2\hbar^2\kappa}\right] \label{eq:axion_field_action}\, ,
\end{align}
where $R$ is the Ricci scalar, $g_{\mu\nu}$ is the spacetime metric tensor (with $(- \: + \: + \: +)$ signature) with determinant $g$ and $\kappa=8 \pi G/c^4$ is the Einstein gravitational constant.
At cosmological scales, when $m_a c^2/\hbar \gg H$, with $H$ the Hubble constant, the axion field equation derived from Eq.~\eqref{eq:axion_field_action} admits  plane wave solutions, i.e.\footnote{In reality, galactic DM is a stochastic superposition of plane waves (see e.g. \cite{foster:2018aa}) that has a finite coherence time. Therefore, the observables of interest in this paper (Eqs.~\eqref{eq:delta_c}, \eqref{eq:Doppler_axion_photon} and \eqref{eq:Sagnac_combination_signal}) will not be purely monochromatic. For the frequencies of interest here, this stochastic feature will not change significantly the sensitivity estimates.}
\begin{subequations}
\begin{align}\label{eq:axion_plane_wave}
    a(t, \vec x) &= a_0 \cos(\omega_a t - \vec k_a \cdot \vec x + \Phi) \, ,
\end{align}
where $a_0, \Phi$ are respectively the dimensionless amplitude and an unknown phase, and $|\vec k_a| = \sqrt{\omega^2_a/c^2-m^2_a c^2/\hbar^2}$ is the wavevector of the field. If the axion field is identified as the nonrelativistic DM, one has
\begin{align}
    a_0 &= \frac{\sqrt{16 \pi G \rho_\mathrm{DM}}}{\omega_a c}\, \\
    \vec k_a &= -\frac{\omega_a v_\mathrm{DM}}{c^2}\hat e_v\, \label{eq:k_a},
\end{align}
where $\rho_\mathrm{DM} = 0.4$ GeV/cm$^3$ is the local DM energy density \cite{McMillan11}\footnote{The actual numerical value of $\rho_\mathrm{DM}$ depends on the model used to describe the velocity distribution of stars around the galaxy and can vary from $\sim 0.3$ GeV/cm$^3$ to $\sim 0.7$ GeV/cm$^3$, see \cite{Cirelli24,Evans19}. This is why most DM sensitivity plots (like the one we show in Fig.~\ref{fig:axion_photon_GW}) assume a specific value of $\rho_\mathrm{DM}$ for consistent comparison.}, $\vec v_\mathrm{DM} = v_\mathrm{DM}\hat e_v \sim 10^{-3}\: c \hat e_v$ is the mean velocity of the Sun in the galactic halo and 
\begin{equation}
    \omega_a \sim \frac{m_a c^2}{\hbar}
\end{equation}
(at leading order in $(v_\mathrm{DM}/c)^2$).
\end{subequations}
As we shall see in the following, we will be interested in experiments with typical length scales corresponding to the distance between the S/C exchanging light signals $L \leq 3 \times 10^9$ m, sensitive to frequencies in the band $f_a \in [10^{-4},10]$ Hz. Thus, the de Broglie wavelength of the field exceeds largely the distance between the S/C $\lambda_a = 2 \pi/|\vec k_a| \geq 3 \times 10^{10} \: \mathrm{m}\gg L$, or in other words, the axion field is highly homogeneous on the scale of the experiment. Therefore,  the propagation phase of the plane wave can be neglected such that Eq.~\eqref{eq:axion_plane_wave} becomes
\begin{align}\label{eq:axion_plane_wave_2}
    a(t, \vec x) &\approx \frac{\sqrt{16 \pi G \rho_\mathrm{DM}}}{\omega_a c^2} \cos(\omega_a t + \Phi) \, ,
\end{align}

In addition to the action Eq.~\eqref{eq:axion_field_action}, we add the Lagrangian describing the interaction in vacuum between the axion $a$ and EM through the coupling $g_{a\gamma}$ \cite{Sikivie21}
\begin{align}\label{eq:axion_photon_lagrangian}
\mathcal{L} &= -\frac{1}{4\mu_0}F_{\mu\nu} F^{\mu\nu} - E_P\frac{g_{a\gamma}}{4\mu_0}aF_{\mu\nu}\tilde F^{\mu\nu}\, , 
\end{align}
where the EM gauge field $A^\alpha$ has usual units of V.s/m, where $F^{\mu\nu}$ is the EM strength tensor, $\tilde F^{\mu\nu} = 1/2\epsilon^{\mu\nu\rho\sigma}F_{\rho\sigma} \equiv \epsilon^{\mu\nu\rho\sigma} \partial_\rho A_\sigma $ is the dual EM strength tensor, with $\epsilon^{\mu\nu\rho\sigma}$ the antisymmetric 4-dimensional Levi-Civita symbol, $\mu_0$ is the vacuum permeability and $E_P= \hbar c /\kappa \sim 10^{18}$ GeV is the reduced Planck energy.

From Eq.~\eqref{eq:axion_photon_lagrangian}, we can derive the EM field equations
\begin{align}
\label{EOM_EM}
\square{A^\nu} - \partial^\nu \partial_\mu A^\mu+ E_Pg_{a\gamma}(\partial_\mu a)\epsilon^{\mu\nu\rho\sigma} \partial_\rho A_\sigma &= 0 \, .
\end{align}
We now focus on the spatial equation of motion of the EM field (i.e $\nu \neq 0$ in Eq.~\eqref{EOM_EM}). Assuming both temporal and Coulomb gauges for the EM field $A_0=0$ and $\vec{\nabla} \cdot \vec{A} = 0$, one finds \cite{Obata18}
\begin{align}\label{eq:EoM}
\ddot{A_i} - c^2 \nabla^2 A_i + cE_P g_{a\gamma}\dot{a}\epsilon_{ijk}\partial_j A_k = 0 \, .
\end{align}
Decomposing light in the two right $A^+$ and left $A^-$ circular polarizations, Eq.~\eqref{eq:EoM} becomes \cite{Obata18}
\begin{subequations}\label{eq:EoM_left_right_polar}
\begin{align}
&\ddot{A}^+_i = - k^2 c^2\left(1 + \frac{\sqrt{16\pi G \rho_\mathrm{DM}}g_{a\gamma}E_P}{kc^2} \sin(\omega_a t +\Phi)\right)A^+_i \, \\
&\ddot{A}^-_i= -k^2 c^2 \left(1 - \frac{\sqrt{16\pi G \rho_\mathrm{DM}}g_{a\gamma}E_P}{k c^2} \sin(\omega_a t +\Phi)\right)A^-_i \, ,
\end{align}
\end{subequations}
respectively for the right and left polarizations.
With a plane wave Ansatz, one finds that the left and right modes have different dispersion relations, i.e both circular polarizations travel with different phase velocities $c_\pm = \omega_\pm/k$, respectively \cite{Obata18}
\begin{subequations}
\begin{align}
\label{delta_c_axion_photon}
c_\pm &= c\sqrt{1 \pm \frac{\sqrt{16\pi G \rho_\mathrm{DM}}g_{a\gamma}E_P}{k c^2} \sin(\omega_a t +\Phi)} \\
 &\equiv c  \pm \delta c(t) \, ,
\end{align}
with
\begin{equation}\label{eq:delta_c}
    \delta c(t)\approx \frac{\sqrt{4\pi G \rho_\mathrm{DM}}g_{a\gamma}}{kc^2} \sin(\omega_a t +\Phi) \, .
\end{equation}
\end{subequations}

In other words, vacuum becomes birefringent in presence of an axion background. Experiments such as DANCE \cite{Obata18,Michimura20} attempt to detect the axion-photon coupling through this effect. We will now show how space-based GW detectors could probe this coupling via a slight modification of their OB. 

\section{Optical benches}\label{sec:OB}

\begin{figure}
    \centering
    \includegraphics[width=0.25\textwidth]{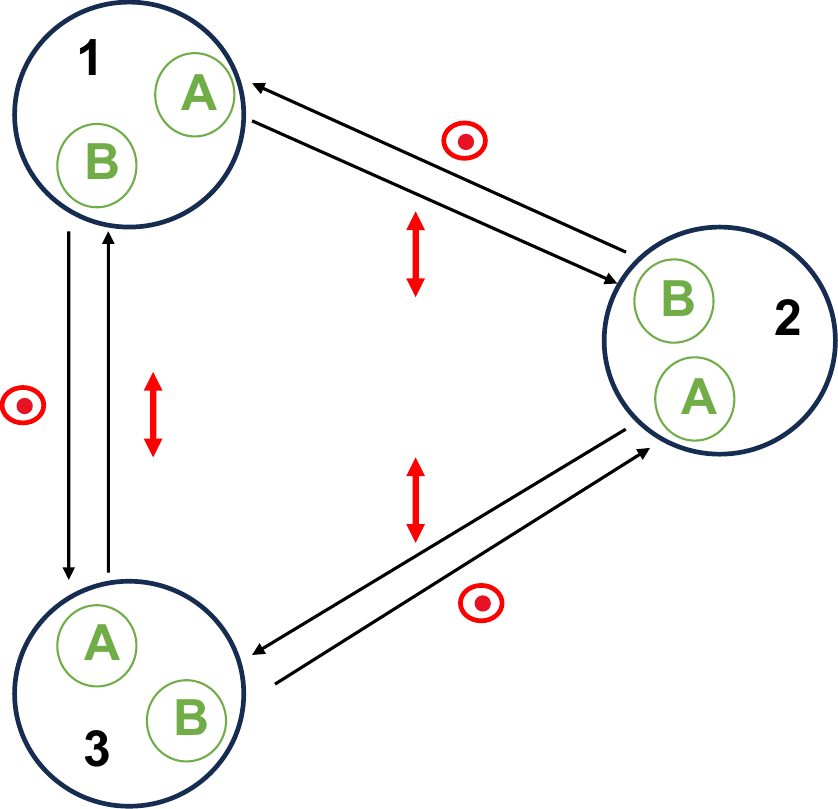}
    \caption{The constellation involves three S/C, denoted $1,2,3$, each containing two optical benches as described in Fig.~\ref{fig:opti_bench_LISA} denoted by "A" and "B". Six different light signals are sent between the different S/C to form the interferometer; three vertically polarized beams are propagating clockwise and the other three horizontally polarized beams are propagating counterclockwise.}
    \label{fig:LISA_const}
\end{figure}

In this section, we will be interested in the polarization of light inside the detectors OB and in free space, between the S/C. As a reminder, all the experiments under consideration in this paper will consist of one (or more) constellation of three S/C, in the following noted $1, 2, 3$.
Each S/C is hosting two OB (labeled $A$ and $B$) with one laser per OB. These two laser beams are sent to the two other S/C and GW signals will be detected by interferometric measurement between those optical signals (see Fig.~\ref{fig:LISA_const}). For our purposes, we are interested in the polarization of these  beams  inside the S/C and also in  between the S/C. 
\subsection{Current design for the optical benches}

The current (simplified) design for the OB envisioned for space-based GW observatory is shown in Fig.~\ref{fig:opti_bench_LISA}, from \cite{Bayle23,Luo16,Luo20}, where we only take into account the interference between laser beams of the local S/C with a distant one. We assume for simplicity that the light wave is produced with a linear polarization along the $x$-axis, is travelling along the $z$ direction and arrives on the detector with a polarization along the $y$-axis after passing through the half-wave plate. 

\subsection{Requirements for GW detection} \label{sec:req_GW}

\begin{figure}
    \centering
    \includegraphics[width=0.4\textwidth]{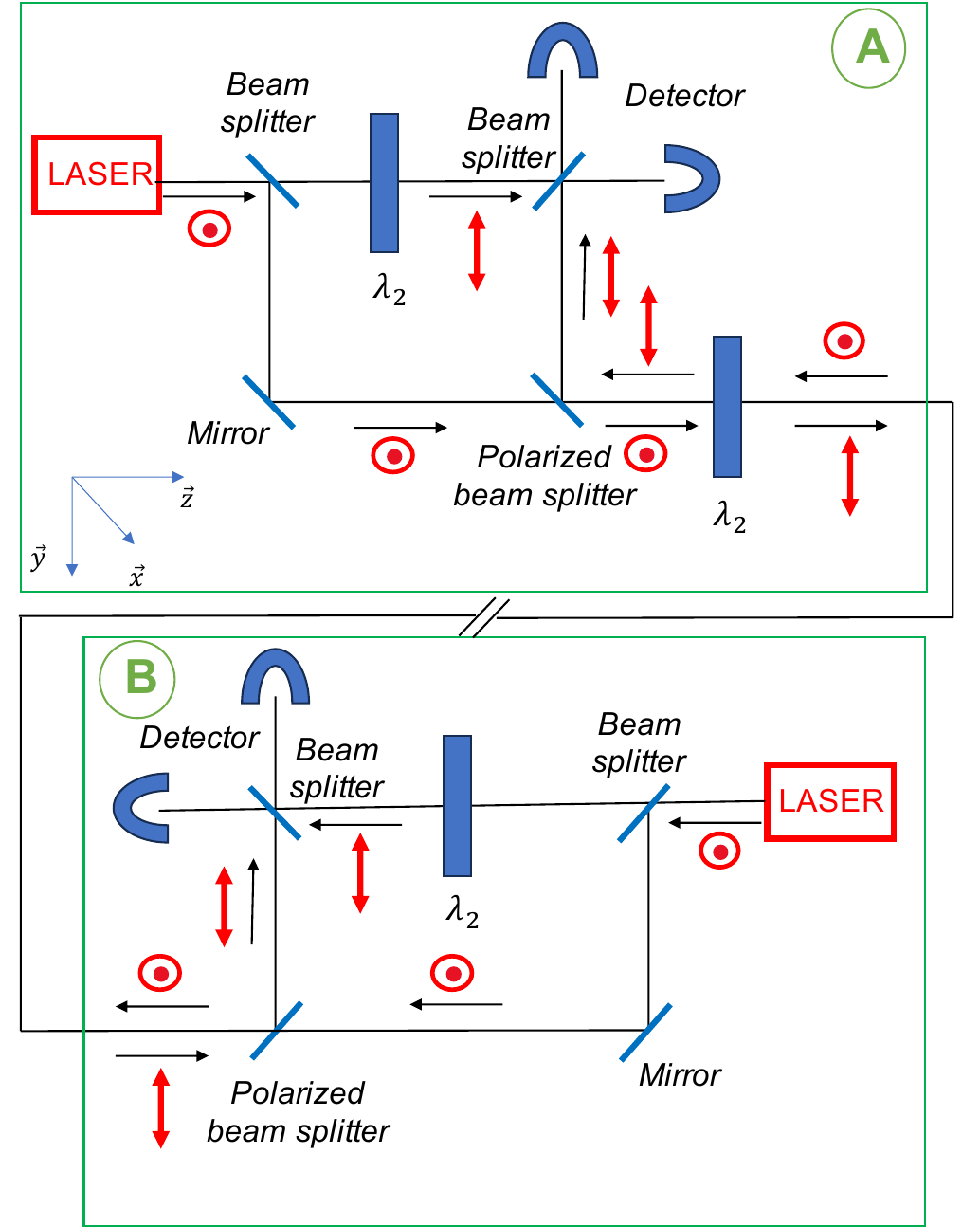}
    \caption{Simplified OB design currently envisioned for space-based GW detectors. The two OB $A$ and $B$ are inside two different S/C (shown as the light green cube) and exchange laser beams between each other. The optical elements are shown in blue, the light polarization in red and the direction of propagation with black arrows. Inside each of these OB, an initial linear polarization directed along the $x$-axis is produced by the laser and the association of the optical elements inside each bench fulfill all the necessary requirements to enable GW detection (see text).}
    \label{fig:opti_bench_LISA}
\end{figure}

In terms of light polarization, several requirements are necessary for the detection of GW in space-based detectors, as follows
\begin{itemize}
    \item In each OB, the incoming beam (i.e the one which has travelled through free space from the distant S/C) must be, at least partially, reflected by the polarized beam splitter to enter the interferometer and interfere with the local beam.
    % \item The two beams interfering inside each OB (one from the local laser and one from the distant S/C, which has travelling in free space) must not have orthogonal polarizations, otherwise the interference between the two beams is zero.
    \item When leaving one S/C, a small portion of the local beam will be undesirably reflected on it, coming back inside the OB to create an additional noise in the interferometer\footnote{Assuming the frequency of the local beam to be $f_\ell$, the distance beam has a slightly different frequency $f_\ell + \delta f$, due to standard Doppler from S/C differential velocity, and frequency planning, see \cite{Heinzel24}. The reflected component of the local beam could interfere with itself but would only create a DC component. However, it could also interfere with the distant beam, and create a spurious beatnote at $\delta f$. In addition, the motion of the reflecting component at the edge of the S/C would induce a time dependent phase on the reflected beam, which would mimic a signal from the GW, and greatly alter the measurement.}. We require this reflected component to be rejected by the polarized beam splitter, such that it does not enter the interferometer. 
\end{itemize}
% Each S/C contains two OB, as shown in Fig.~\ref{fig:opti_bench_LISA}. Inside one given OB (say $A$ in S/C $1$), two beams are interfering : the one produced by the local laser of the given OB and the one which has been produced by one of the OB of another S/C ($B$ in S/C $2$) and which has been travelling in free space. We require the two beams to have the same polarization, when they enter the detectors inside OB $A$ (or at least that they do not have orthogonal polarizations), otherwise the interference between the two beams is zero (this requirement is the same in OB $B$).

% When leaving one S/C, a small portion of the local beam will be undesirably reflected on it, come back inside the OB and create an additional noise in the interferometer. 
% To overcome this issue, we require the outgoing polarization to be orthogonal to the incoming one. Then, the polarized beam splitter selects only the polarization of interest, i.e the one of the incoming beam.

It can be shown easily that the current OB design from Fig.~\ref{fig:opti_bench_LISA} fulfills those requirements\footnote{There are likely several additional effects that need to be carefully studied before an implementation of the modification that we propose (e.g. additional stray light from the polarizer itself, angular or translational motion of the beam induced by the polarizer, phase fluctuations induced by the polarizer, etc…). Evaluating them in detail is beyond the scope of this paper, but we note that those effects are similar to the “usual” effects already studied and under control for other optical components, like the $\lambda/2$ wave plates in the original design. We therefore assume that suitable solutions can be found, and restrain this paper to introducing the principle and estimating the corresponding sensitivity.}.
In the full constellation (with the three S/C) with two OB in each S/C, six individual interference measurements are performed at each time. Two different linear polarizations travel in space : vertical polarization travels in one direction (say clockwise) while horizontal polarization travels in the other direction (say counterclockwise) (see Fig.~\ref{fig:LISA_const}).

\subsection{Requirements for axion detection}

In addition to the previous requirements for space-based GW detectors to work, the additional condition necessary for the axion-photon detection is
\begin{itemize}
    \item The two beams travelling in free space between the S/C (e.g. $A \rightarrow B$ and $B \rightarrow A$) must have a different polarization, and at least one of them has to be circularly polarized. 
\end{itemize}
Indeed, as derived in Section.~\ref{sec:vacuum_birefringence}, the axion field only impacts the phase velocity of circularly polarized light and not linearly polarized one\footnote{In reality, the coupling also affects the two linear polarization states because it couples them, i.e a linear state slightly oscillates into the other one and vice-versa. Using e.g. an initial pure "horizontal" polarization state, one could see such effect using a vertical polarizer and counting the photons that go through due to the rotation of the polarization plane. This is essentially the detection idea in e.g. \cite{Nagano19}.}. Therefore, in order to detect the axion field one must introduce other optical elements inside the original OB shown in Fig.~\ref{fig:opti_bench_LISA} in order to generate circularly polarized light, whilst still satisfying the requirements for GW detection of Sect. \ref{sec:req_GW}.  
\begin{figure}
    \centering
    \includegraphics[width=0.5\textwidth]{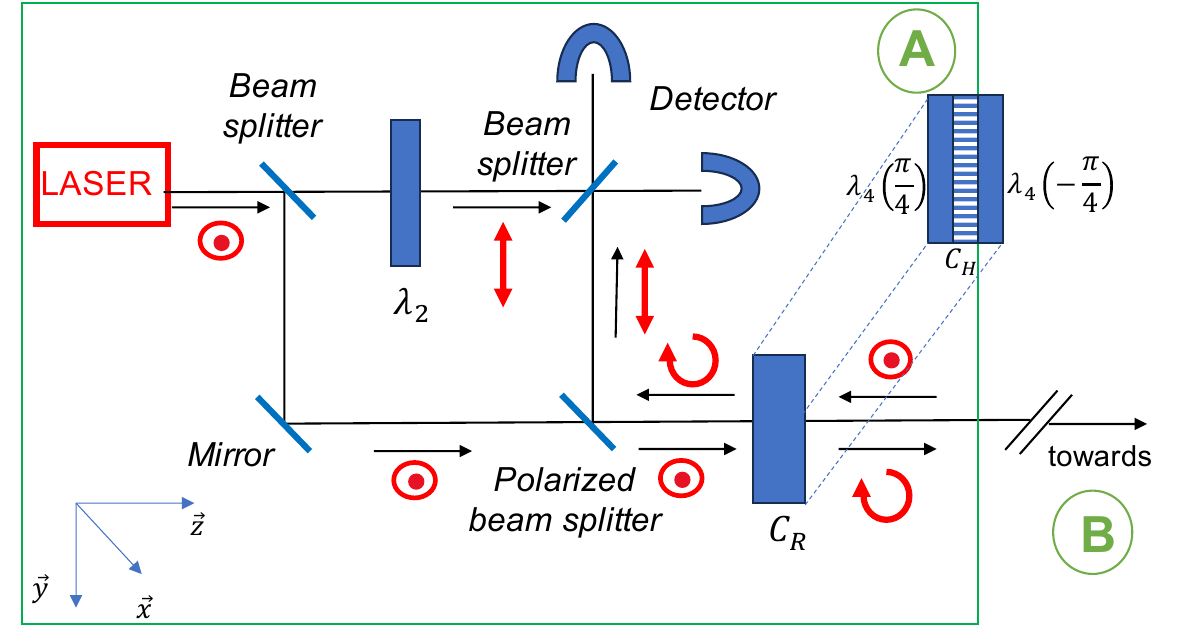}
    \caption{Modified OB design for the OB $A$ (to be compared to the original one presented in Fig.~\ref{fig:opti_bench_LISA}), in order to be sensitive to vacuum birefringence induced by the axion-photon coupling. The half-wave plate at the end of the OB $A$ is replaced by a right polarizer which is a combination of a quarter-wave plate, a horizontal polarizer and a second quarter-wave plate, are oriented conveniently. This configuration produces a right polarized light at the output of OB $A$, and fulfill all the requirements for the optical interferometer to work. The OB $B$ is unchanged.}
    \label{fig:opti_bench_LISA_modified}
\end{figure}

\subsection{Proposed modification}\label{sec:solution}

\begin{figure}[b!]
\centering
\includegraphics[width=0.25\textwidth]{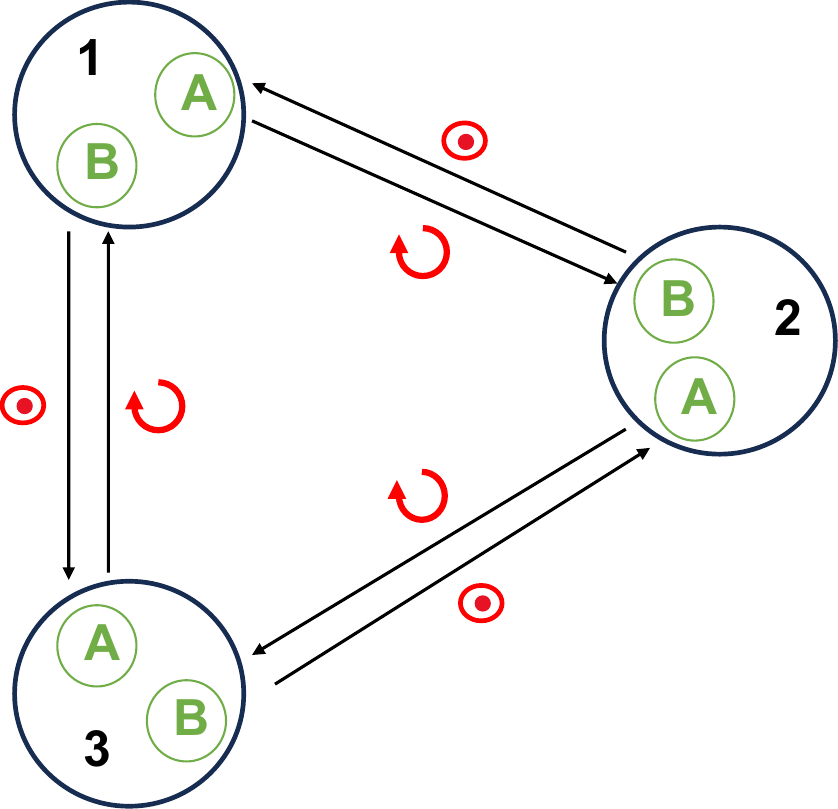}
\caption{Proposed modified constellation for the search of vacuum birefringence, as a consequence of Fig.~\ref{fig:opti_bench_LISA_modified}. In this framework, only the arms where light is circularly polarized are impacted by the axion field, and get phase shifted (see text).}
\label{fig:LISA_const_modified}
\end{figure}

A solution to all the requirements presented in the last two sections is to replace the last $\hat \lambda_2$ in the OB $A$ (see Fig.~\ref{fig:opti_bench_LISA}), by a $\hat \lambda_4(\pi/4)\hat C_H \hat \lambda_4(-\pi/4)$ combination, i.e a first quarter-wave plate oriented with $\theta=\pi/4$, then a horizontal linear polarizer and then a second quarter wave plate oriented with $\theta=-\pi/4$, as shown in Fig.~\ref{fig:opti_bench_LISA_modified}. This corresponds to a right polarizer. 
In Appendix.~\ref{ap:jones_solution}, we show mathematically how this solves the various requirements, using the Jones formalism.
Therefore, through this modification of the OB, half of the beams are circularly polarized (see the modified full constellation in Fig.~\ref{fig:LISA_const_modified}), and therefore the detection of the axion-photon coupling would be possible.

\section{Sensitivity of space-based GW observatory to the axion-photon coupling}\label{sec:results}
In the previous section, we have demonstrated that a minor change in the OB design from space-based GW observatories can make them sensitive to the coupling between an axion and electromagnetism. Let us now assess quantitatively the sensitivity of these space missions to $g_{a\gamma}$.

\subsection{Single-arm Doppler effect}

In this section, we derive the single-arm Doppler effect in the GW space-based detectors induced by the axion-photon coupling, in the case where the OB modification shown in Fig.~\ref{fig:opti_bench_LISA_modified} is made. As a reminder, this coupling modifies the phase velocity of left and right circularly polarized light as derived in Eq.~\eqref{delta_c_axion_photon}, while the phase velocity of linearly polarized light is unchanged.

In Appendix.~\ref{ap:phase_shift}, we derive the expected phase shift of light induced by vacuum birefringence on the detectors arms. It reads 
\begin{align}\label{eq:phase_shift_axion_photon_LISA}
\Delta \phi(t) &= \frac{\sqrt{16\pi G \rho_\mathrm{DM}} E_P g_{a\gamma}}{\omega_a c} \sin\left(\frac{\omega_a L}{2c}\right)\sin\left(\omega_a t+\Phi'\right) \,\\
&\equiv \Delta \phi_\mathrm{single \: link} \, \nonumber ,
\end{align}
for a right handed polarized EM wave where $\Phi' = \Phi - \omega_a L/2c$. Therefore, following Fig.~\ref{fig:LISA_const_modified}, we have 
\begin{subequations}\label{eq:phase_shift_arms}
\begin{align}
    \Delta \phi_{12} &= \Delta \phi_{23} = \Delta \phi_{31} = 0 \, \\
    \Delta \phi_{13} &= \Delta \phi_{32} = \Delta \phi_{21} = \Delta \phi_\mathrm{single \: link} \, ,
\end{align}
\end{subequations}
where the $ij$ subscript denotes the arm pointing from S/C $j$ to S/C $i$. We define the relative frequency fluctuation signals (i.e Doppler shifts) as 
\begin{align}\label{eq:Doppler_phase_shift_link}
    y_{ij}(t) &= \frac{1}{\omega_j}\frac{d \Delta \phi_{ij}}{dt} \, , 
\end{align}
where $\omega_j$ is the local laser frequency in S/C $j$. Using Eqs.~\eqref{eq:phase_shift_axion_photon_LISA}, \eqref{eq:phase_shift_arms} and \eqref{eq:Doppler_phase_shift_link}, we find the single-link Doppler shifts as
\begin{subequations}\label{eq:Doppler_axion_photon}
\begin{align}
&y_{12} = y_{23} = y_{31}= 0 \, \\
&y_{13}(t) = y_{32}(t) = y_{21}(t) = \frac{\sqrt{16\pi G \rho_\mathrm{DM}} E_P g_{a\gamma}}{2\pi \nu_0 c} \times \,\nonumber\\
& \sin\left(\frac{\omega_a L}{2c}\right)\cos\left(\omega_a t+\Phi'\right) \, ,
\end{align}
\end{subequations}
where $\omega_j/2\pi \approx \omega_0/2\pi = \nu_0$ is the laser frequency, supposed to be the same inside all OB. As it can be noticed from Eq.~\eqref{eq:Doppler_axion_photon}, the Doppler shift induced by the axion-photon coupling is unpolarized in the sense that it does not depend on  e.g. the polarization of the incoming wave (as it is the case for the GW, see e.g. \cite{Cornish03}) or the propagation direction of the incoming wave (as it is the case for other DM couplings, see e.g. \cite{Yu23,Morisaki19}). Therefore, this signal differs significantly from other signals expected to be detected in the space missions of interest in this paper.

\subsection{Sagnac $\alpha$ TDI combination}

In space-based GW detectors, the laser noise on single link measurement is expected to dominate largely the signal by at least 8 orders of magnitude, which would make the GW detection impossible.
Fortunately, there exists a method to effectively reduce drastically the laser noise by combining signals from the different S/C in such a way that the resulting observable is nearly independent of laser noise, but still contains the GW signal (or the DM signal in our case). This is called \textit{Time Delay Interferometry}, or TDI and it was first introduced in \cite{Armstrong99, Tinto99, Dhurandhar02}, for a detailed review see \cite{tinto:2021aa}. The principle relies on the fact that each laser noise is measured several times. Some specific time retarded signal combinations between the six different arms strongly reduces the laser noise to zero, thus the name of the method. In this section, we will only be interested in one specific TDI combination, the Sagnac $\alpha$, which is the most effective TDI combination to use for the axion-photon detection. Indeed, the Sagnac $\alpha$ measures the phase difference of two virtual laser beams starting at the same position and going around the constellation, one clockwise and the other counterclockwise. In our case, as Fig.~\ref{fig:LISA_const_modified} suggests, one of the two contributions (the one travelling clockwise) would get maximized, while the other one (the links travelling counterclockwise) would not contribute, and therefore the signal is maximum. In the following, we show how the one-link signal Eq.~\eqref{eq:Doppler_axion_photon} can get amplified through the use of the Sagnac $\alpha$ combination.

In the time domain, the second generation Sagnac $\alpha$ combination is defined as (see the definition of $C_1^{12}$ in Table 1 from \cite{hartwig:2022aa})
\begin{subequations}
\begin{align}
    %\alpha(t) &= y_{12}+D_{12}y_{23}+D_{123}y_{31}\nonumber\\
    %&+D_{1231}y_{13} + D_{12313} y_{32} + D_{123132} y_{21} \nonumber \\
    %&-y_{13}-D_{13}y_{32}-D_{132}y_{21}\nonumber\\
    %&-D_{1321}y_{12}-D_{13212}y_{23}
    \alpha(t)& = \left(1-D_{1321}\right)\left(y_{12}+D_{12}y_{23}+D_{123}y_{31}\right)\nonumber\\
    &-\left(1-D_{1231}\right)\left(y_{13}+D_{13}y_{32}+D_{132}y_{21}\right)\, \label{eq:Sagnac_combination},
\end{align}
where we use the compacted writing of TDI $D_{ij}y_{jk}(t) = y_{jk}\left(t-L_{ij}(t)/c\right)$ and $D_{ijk}y_{kl}(t) = y_{kl}\left(t-L_{ij}(t)/c-L_{jk}\left(t-L_{ij}(t)/c\right)\right)$. Using Eq.~\eqref{eq:Doppler_axion_photon} and the constant and equal arm-length approximation (i.e. $L_{ij}(t)=L$), Eq.~\eqref{eq:Sagnac_combination} becomes
\begin{align}
    &\alpha(t) = \frac{\sqrt{16\pi G \rho_\mathrm{DM}} E_P g_{a\gamma}}{\pi \nu_0 c}\sin^2\left(\frac{3\omega_a L}{2c}\right)\sin(\omega_a t + \Phi'') \,  , \label{eq:Sagnac_combination_signal}
\end{align}
\end{subequations}
with $\Phi''=\Phi' - 5\omega_a L/2c = \Phi - 3\omega_a L/c$. While the signal has been derived assuming constant and equal arm-lengths, real S/C orbits will induce unequal arm-lengths with time variation at the relative level of the percent (see Fig. 5 from \cite{Martens21}). This will impact the amplitude of the signal from  Eq.~\eqref{eq:Sagnac_combination_signal} (and subsequently the sensitivity on $g_{a\gamma}$) at the percent level at maximum for axion masses below $10^{-15}$ eV. At higher masses, only the exact location of the resonances will be impacted at the percent level.
The one-sided power spectral density (PSD) of this signal reads
\begin{align}
    S_\alpha(f) =& \left(\frac{\sqrt{16\pi G \rho_\mathrm{DM}} E_P g_{a\gamma}}{\pi \nu_0 c}\right)^2\sin^4\left(\frac{3\omega_a L}{2c}\right)\,\nonumber\\
    & \quad \times \frac{1}{2\pi^2 T_\mathrm{obs}}\left(\frac{\sin (\pi(f-f_a)T_\mathrm{obs})}{f-f_a}\right)^2 \, \label{eq:signal_PSD} ,
\end{align}
where $f$ is the Fourier frequency, $f_a \equiv \omega_a/(2\pi)$ and $T_\mathrm{obs}$ is the total duration of the experiment. For simplicity we have used a square window of duration $T_\mathrm{obs}$ to calculate the PSD, which is sufficient for our rough sensitivity estimate.

The noise PSD of the $\alpha$ TDI channel depends mainly on two sources: (i) on the single-link optical metrology system (OMS) noise, which corresponds to the noise on the optical readout of the interference signals, and (ii) the single test mass acceleration noise, which corresponds to the noise coming from the inertial system comprising the test masses and all elements interacting directly with them. The shape and amplitude of those noises depend on the experiment under consideration. As a reminder, in this paper, we consider four space-based GW detectors: LISA, TianQin, Taiji and BBO. The first three of them consist in an individual triangular constellation either in a heliocentric orbit for LISA \cite{Colpi24} and Taiji \cite{Luo20} or geocentric one for TianQin \cite{Luo16}. BBO will operate four different triangular constellations, two of them located at almost the same position and the two other at different locations of the same heliocentric orbit \cite{Crowder05}.

In their current design, the PSD of the OMS and acceleration noises for LISA, TianQin and Taji are provided by
\begin{subequations}
\begin{align}\label{eq:noise_PSD_LISA}
S^i_\mathrm{oms}(f) &= \left(P^i_\mathrm{oms}\frac{2\pi f}{c}\right)^2\left(1+\left(\frac{2 \: \mathrm{mHz}}{f}\right)^4\right) \: \mathrm{Hz}^{-1} \, \\
S^i_\mathrm{acc}(f) &= \left(\frac{P^i_\mathrm{acc}}{2\pi f c}\right)^2\left(1+\left(\frac{0.4 \: \mathrm{mHz}}{f}\right)^2\right)\,\nonumber \\
&\left(1+\left(\frac{f}{8 \: \mathrm{mHz}}\right)^4\right)\: \mathrm{Hz}^{-1} \, ,
\end{align}
for $i=\{$LISA, TianQin and Taiji$\}$, with \cite{Robson19,Babak21,Luo16,Hu17} 
\begin{align}
    &P^\mathrm{LISA}_\mathrm{oms} = 15 \times 10^{-12} \: \mathrm{m} \: ,  \: P^\mathrm{LISA}_\mathrm{acc} = 3 \times 10^{-15} \: \mathrm{m/s}^2 \, \\
    &P^\mathrm{TianQin}_\mathrm{oms} = 10^{-12} \: \mathrm{m} \: , \: P^\mathrm{TianQin}_\mathrm{acc} = 10^{-15} \: \mathrm{m/s}^2 \, \\
   &P^\mathrm{Taiji}_\mathrm{oms} = 8 \times 10^{-12} \: \mathrm{m} \: , \: P^\mathrm{Taiji}_\mathrm{acc} = 3 \times 10^{-15} \: \mathrm{m/s}^2 \, ,
\end{align}
while for BBO the single-link noises are given by \cite{Corbin06}
\begin{align}
S^\mathrm{BBO}_\mathrm{oms}(f) &= \frac{1.0 \times 10^{-34}}{(3L^2)} \: \mathrm{m}^{2}/\mathrm{Hz} \, \\
S^\mathrm{BBO}_\mathrm{acc}(f) &= \frac{4.5 \times 10^{-34}}{(2\pi f)^4(3L)^2}\:  \mathrm{m}^{2}.\mathrm{Hz}^{4}/\mathrm{Hz} \, ,
\end{align}
\end{subequations} 
where we added a factor $1/2$ to both contributions, to take into account the fact that the two triangular detectors located at the same position will probe the same amplitudes and phases of the field\footnote{The two other detectors are much further away, and their distance is beyond one coherence length for frequencies $f \gtrapprox 1$ Hz, such that they will probe a different signal. Therefore we ignore them for our rough sensitivity estimate.}.

The modification of the OB suggested in Fig.~\ref{fig:opti_bench_LISA_modified} implies a decrease in the amplitude of light (or equivalently in the number of photons $N_\gamma$) that reaches each interferometer by a factor $2$ because light gets partially absorbed by the right polarizer (a linear polarization being a superposition of left and right states, the right polarizer transmits only half of the light intensity from the linear input). Since the OMS noise includes a contribution from shot noise whose PSD $\propto 1/N_\gamma$, we will conservatively add a factor $2$ to the whole OMS contribution\footnote{The sensitivity to GW strain scaling as the square root of the TDI noise PSD, this implies a factor $\sqrt{2}$ loss of sensitivity on $h$.}, such that the Sagnac $\alpha$ noise PSD reads 
\cite{Hartwig23}\footnote{In \cite{Hartwig23}, only the first generation noise PSD are presented (see their Eq. (B2a)). In order to compute the second generation noise PSD for the Sagnac $\alpha$ combination, one needs to add an additional factor $(2\sin(3\omega L/2c))^2$.}
\begin{align}\label{eq:TDI_Sagnac_noise}
    N_\alpha(f) &= 4\sin^2\left(\frac{3 \pi f L}{c}\right)\Bigg[12S_\mathrm{oms}(f) +\\
    &4\left(3-2\cos\left(\frac{2\pi f L}{c}\right)-\cos\left(\frac{6\pi f L}{c}\right)\right)S_\mathrm{acc}(f)\Bigg] \, \nonumber .
\end{align}

\subsection{Expected sensitivity}

We now estimate the sensitivity to $g_{a\gamma}$ of each experiment under consideration in this paper.
Since we assume a signal-to-noise ratio (SNR) of $1$, a simple (and conservative) estimate of the detectable coupling is obtained by equating the signal PSD $S_\alpha(f)$ (given by Eq.~\eqref{eq:signal_PSD} in the limit $f\to f_a$) to the noise PSD $N_\alpha(f)$ (given by Eq.~\eqref{eq:TDI_Sagnac_noise}), i.e.
\begin{subequations}
\begin{align}\label{eq:sens_mono_plane_wave}
    g_{a\gamma}(f) &= \frac{\pi \nu_0 c}{\sqrt{8\pi G \rho_\mathrm{DM}} E_P\sin^2\left(\frac{3 \pi f L}{c}\right)}\sqrt{\frac{N_\alpha(f)}{T_\mathrm{obs}}}\, .
\end{align}
Galactic DM is not purely monochromatic, it is a superposition of plane waves which oscillate at slightly different frequencies (see e.g. \cite{Derevianko18, foster:2018aa}). Therefore, the field acquires a characteristic coherence time $\tau(f) \sim 10^6/f$ \cite{Derevianko18}. Eq.~\eqref{eq:sens_mono_plane_wave} is valid when $T_\mathrm{obs} \ll \tau(f)$, i.e when the field can effectively be described as a monochromatic plane wave.
Note that a correction factor to the sensitivity arises due to the stochastic nature of the amplitude of the field \cite{Centers21} which induces a loss in signal of $\sim 1.5$ for a SNR of 1.
When the integration time is much longer than the coherence time of the field, i.e $T_\mathrm{obs} \gg \tau(f)$, the signal searched for is no longer coherent, i.e. it should be modeled as a sum of several stochastic harmonics, see \cite{foster:2018aa}, and therefore Eq.~\eqref{eq:signal_PSD} is not valid anymore. Another method to analyse the data is to cut the dataset in fragments with duration smaller than $\tau(f)$ and search for a coherent signal in each of these blocks of data. In such a case, the experimental sensitivity to the coupling is reduced and becomes \cite{Budker14}
\begin{align}
    g_{a\gamma}(f) &= \frac{\pi \nu_0 c }{\sqrt{8\pi G \rho_\mathrm{DM}} E_P\sin^2\left(\frac{3 \pi f L}{c}\right)}\sqrt{\frac{N_\alpha(f)}{\sqrt{T_\mathrm{obs}\tau(f)}}} \, .
\end{align}
\end{subequations}

\begin{table*}[t!]
\centering
\begin{tblr}{
    vlines,
    colspec={ccccc}
}
\hline
& LISA & TianQin  & Taiji & BBO\\
\hline\hline
Laser frequency $\nu_0$ (Hz) & $2.82 \times 10^{14}$ & $2.82 \times 10^{14}$ & $2.82 \times 10^{14}$& $6 \times 10^{14}$ \\
\hline
Arm length $L$ (m) & $2.5 \times 10^9$ & $\sqrt{3} \times 10^8$ & $3 \times 10^9$& $5 \times 10^7$ \\
\hline
Integration time $T_\mathrm{obs}$ (year) & $4.5$ & $5 \times 1/2$ & $5$ & $4$ \\
\hline
Frequency band (Hz) & $[10^{-4}-1]$ & $[10^{-4}-1]$ & $[10^{-4}-1]$& $[10^{-1}-10]$ \\
\hline
\end{tblr}
\caption{Experimental parameters of interest for LISA, TianQin, Taiji and BBO, from \cite{Colpi24,Nam23,Luo16,Luo20,Corbin06,Cutler09}. The time of integration of TianQin suffers from a $1/2$ factor because the detector plane is fixed throughout the orbit, implying it will be blinded by the Sun during half of the orbit. The time of integration of BBO is purely presumed.}
\label{tab:exp_param}
\end{table*}

All experimental parameters of interest for the sensitivity curves are extracted from \cite{Colpi24,Nam23} for LISA, from \cite{Luo16} for TianQin, from \cite{Luo20} for Taiji and from \cite{Corbin06,Cutler09} for BBO. They are summarized in Table ~\ref{tab:exp_param}. 
\begin{figure}
    \centering
    \includegraphics[width=0.5\textwidth]{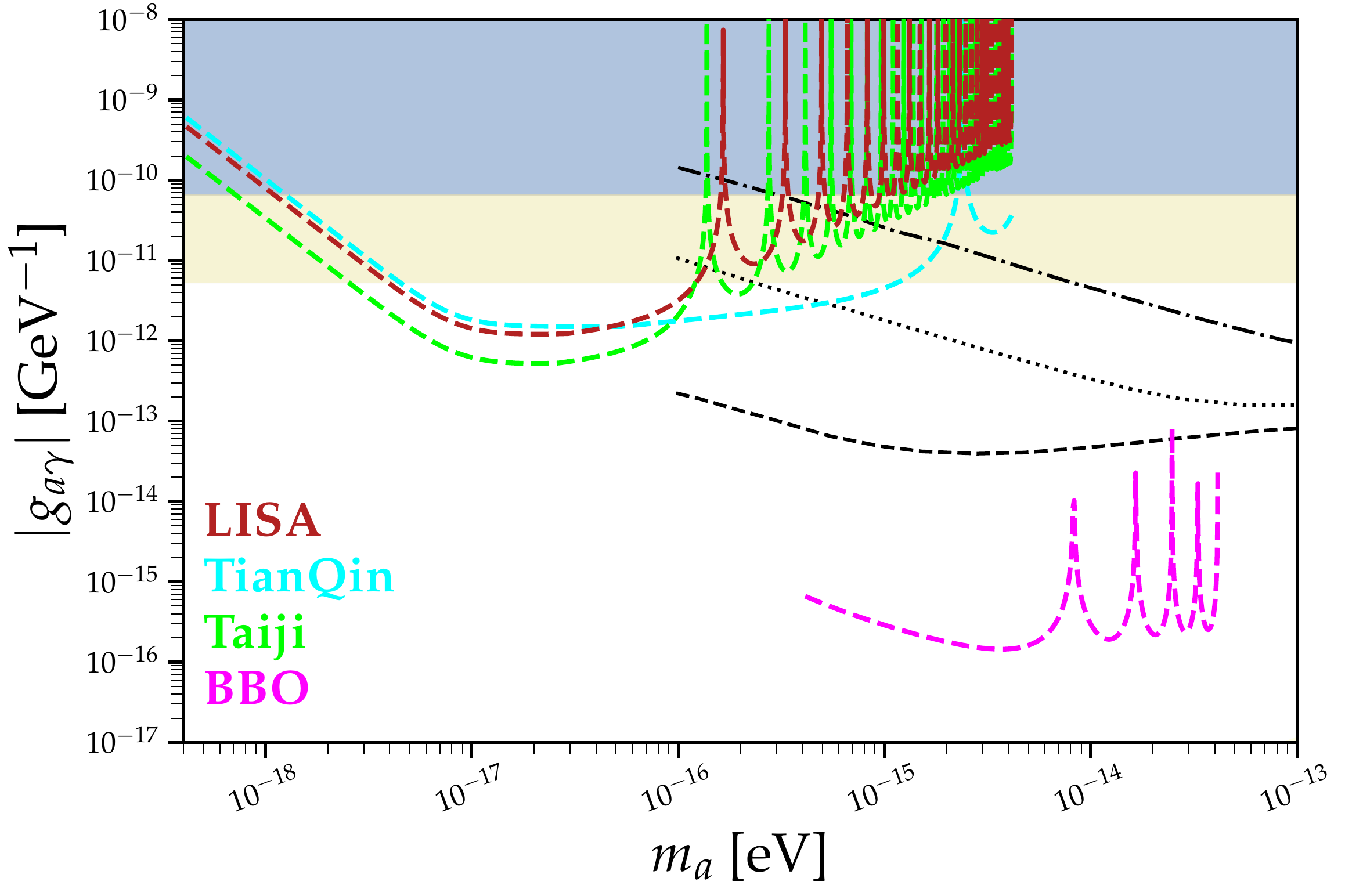}
    \caption{Sensitivity estimates of  various GW space-based detectors (shown in  brown, cyan, lime and magenta respectively for LISA, TianQin, Taiji and BBO) to the axion-photon coupling $g_{a\gamma}$. In light blue and light yellow, we show the respective constraint from CAST \cite{CAST} and observations of SN1987A \cite{Payez15} (from \cite{AxionLimits}). Finally, in dashed-dotted, dotted and dashed black lines, we show respectively the expected sensitivity of advanced LIGO, DECIGO and Cosmic Explorer detectors, from \cite{Nagano19} (which have been rescaled for consistent numerical value of $\rho_\mathrm{DM}$).}
    \label{fig:axion_photon_GW}
\end{figure}
In Fig.~\ref{fig:axion_photon_GW}, we show the respective expected sensitivity of LISA, TianQin, Taiji and BBO, shown in brown, cyan, lime and magenta. One can notice that all these GW space-based detectors would improve the current laboratory constraints on $g_{a\gamma}$, for axion masses between $\sim 4\times 10^{-19}$ to $4 \times 10^{-15}$ eV for LISA, TianQin and Taiji, and between $\sim 4\times 10^{-16}$ to $4 \times 10^{-14}$ eV for BBO. In particular, between $10^{-17}$ and $10^{-15}$ eV, LISA, TianQin and Taiji would respectively reach $g_{a\gamma} \sim 10^{-12}, 1.5 \times 10^{-12}$ and $4 \times 10^{-13}$ GeV$^{-1}$, improving by 1 order of magnitude the current best constraint at these masses, while BBO could reach $g_{a\gamma} \sim 10^{-16}$ GeV$^{-1}$, 5 orders of magnitude better than SN1987A, the current best constraint at these masses.

\section{Discussion}

In this paper, we have shown how operating circular polarized light beams in space-based gravitational wave detectors could transform them into axion detectors, whilst remaining sensitive to GW. Indeed, the coupling between axion and photon with strength $g_{a\gamma}$ induces an oscillation of the phase velocity of circularly polarized light through vacuum birefringence. In space-based GW detectors, it induces a phase shift of circularly polarized light compared to linearly polarized one, which would create an observable Doppler shift. 

This could be made possible by slightly modifying the current OB design, changing one half waveplate by a circular polarizer. The number of photons that arrive at each optical interferometer from the distant S/C would be reduced by a factor 2, which would merely impact the minimum detectable GW amplitude by a small factor $\sqrt{2}$. 

As we have shown it in Fig.~\ref{fig:axion_photon_GW}, operating this new OB would allow the various space-based GW detectors under consideration in this paper to reach unconstrained regions of the $g_{a\gamma}-m_a$ parameter space, improving current best constraints on $g_{a\gamma}$ by as much as 5 orders of magnitude over several orders of magnitude of axion masses. In particular, such detection techniques would be complementary to futuristic versions of GW ground-based detectors, optimized for the axion-photon coupling searches \cite{Nagano19}, which are the most sensitive for axion masses between $\sim 10^{-16}$ and $10^{-11}$ eV (due to the smaller baseline). Note that the modification that we propose would not be optimal for ground based detectors, because they use Fabry-Perot cavities. Indeed, assuming circular states, the polarization of photons change at each reflection inside the cavity. Therefore, the phase accumulated over two reflections will in general cancel (because of the sign difference in Eq.~\eqref{delta_c_axion_photon}), except if the round trip time corresponds exactly to a period of oscillation of the axion field.

\begin{acknowledgments}

This work was supported by the Programme National GRAM of CNRS/INSU with INP and IN2P3 cofunded by CNES. IFAE is partially funded by the CERCA program of the Generalitat de Catalunya. J.G. is funded by the grant CNS2023-143767. Grant CNS2023-143767 funded by MICIU/AEI/10.13039/501100011033 and by
European Union NextGenerationEU/PRTR. 

\end{acknowledgments}

\appendix

\section{OB modification}\label{ap:jones_solution}

In this appendix, we show mathematically how the solution provided in Sec.~\ref{sec:solution} is valid for both GW and axion detection. 

\subsection{Jones formalism}

We can model the OB system using the Jones formalism for polarization states of light and optical elements which is very convenient when treating interference phenomena \cite{Collett05}. This formalism discomposes the light polarization on a basis formed of the two linear polarizations. The polarization state is modelled using a $2\times1$ matrix while the optical elements are represented by $2\times2$ matrices. 

As shown in Fig.~\ref{fig:opti_bench_LISA}, we assume the light beam to propagate along the $z$-axis with an initial polarization along the $x$-axis, that we call "horizontal" polarization. Then, its Jones representation is \cite{Collett05}
\begin{align}
    |H\rangle &= \begin{pmatrix}
        1 \,\\
        0 
    \end{pmatrix} \, .
\end{align}
The other linear polarization along $y$-axis, that we will call "vertical" is represented as 
\begin{align}
    |V\rangle &= \begin{pmatrix}
        0 \,\\
        1
    \end{pmatrix} \, .
\end{align}
A half-wave plate adds a phase $\phi=\pi$ to the light polarization parallel to its slow axis\footnote{The principle of a wave plate is that it presents two perpendicular directions with respective refractive index $n_f, n_s$ with $n_f<n_s$, that we call respectively fast and slow axis \cite{Hecht02}. This leads to a phase shift between the light polarization along the fast axis and the one along the slow axis.}. Its Jones matrix representation is \cite{Collett05}
\begin{align}
    \hat \lambda_2(\theta) &= e^{-i\frac{\pi}{2}}\begin{pmatrix}
        \cos^2 \theta - \sin^2\theta & 2 \sin\theta \cos \theta \,\\
        2 \sin\theta \cos \theta & \sin^2\theta-\cos^2\theta 
    \end{pmatrix} \,\nonumber \\
    &\equiv \begin{pmatrix}
        \cos(2\theta) & \sin(2\theta) \,\\
        \sin(2\theta) & -\cos(2\theta) 
    \end{pmatrix} \, ,
\end{align}
up to an irrelevant total phase, where $\theta$ is the angle between the horizontal axis (the $x$-axis) and the fast axis of the plate. Similarly, a quarter-wave plate introduces a phase shift of $\pi/2$ between the two linear polarized states and can be represented as \cite{Collett05}
\begin{align}
    \hat \lambda_4(\theta) &= \begin{pmatrix}
        \cos^2(\theta)-i\sin^2(\theta)& (1+i)\sin(\theta)\cos(\theta) \,\\
        (1+i)\sin(\theta)\cos(\theta) & \sin^2(\theta)-i\cos^2(\theta) 
    \end{pmatrix}\, ,
\end{align}
up to a total phase. Note that if the half or quarter wave plate is rotated such that its fast or slow axis is along the initial linear horizontal or vertical polarization of light (i.e $\theta=0$ or $\theta=\pi/2$), there is no effect, i.e the input and output polarization are the same (up to an irrelevant phase). Note also that the determinant of these matrices is $1$, i.e they preserve energy, or in other words, they have no absorption.
In the Jones formalism, right and left circular polarizations are represented as \cite{Collett05}
\begin{subequations}
\begin{align}
    |R\rangle &= \frac{1}{\sqrt{2}}\begin{pmatrix}
        1 \,\\
        i 
    \end{pmatrix} \, \\
    |L\rangle &= \frac{1}{\sqrt{2}}\begin{pmatrix}
        1 \,\\
        -i 
    \end{pmatrix} \, .
\end{align}
\end{subequations}
We now introduce horizontal, vertical, left and right polarizers, which respectively transmit only horizontal, vertical, left and right polarizations of the input, defined as \cite{Fymat71}
\begin{subequations}
\begin{align}
    \hat C_H &= \begin{pmatrix}
        1 & 0\,\\
        0 & 0 
    \end{pmatrix} \, \\
    \hat C_V &= \begin{pmatrix}
        0 & 0\,\\
        0 & 1 
    \end{pmatrix} \, \\
    \hat C_L &= \frac{1}{2}\begin{pmatrix}
        1 & i\,\\
        -i & 1 
    \end{pmatrix} \, \\
    \hat C_R &= \frac{1}{2}\begin{pmatrix}
        1 & -i\,\\
        i & 1 
    \end{pmatrix} \, .
\end{align}
\end{subequations}
In practice, the right polarizer is made of the combination $\hat \lambda_4(\pi/4)\hat C_H \hat \lambda_4(-\pi/4)$ (and for the left polarizer, one replaces $\hat C_H$ by $\hat C_V$ or exchange the two quarter-wave plates).
Finally, a mirror reverses the direction of one of the two linear polarizations while letting the other one unchanged. Its representation is \cite{Fymat71}
\begin{align}
    \hat M = \begin{pmatrix}
        -1 & 0 \,\\
        0 & 1 
    \end{pmatrix}\, ,
\end{align}
therefore it is the "horizontal" polarization that gets flipped. 

\subsection{Requirements for GW and axion detection}

Between the polarized beam splitter and the end of the OB, we will model the set of optical elements of $n$ half-wave plates, $m$ quarter-wave plates and $\ell$ polarizers (of any kind) that we add by $\hat S(\theta,\theta')$, where $\theta(\theta')$ is the angle between the fast axis of the half-wave (quarter-wave) plates and the horizontal axis. We will denote this set of elements with superscript $A$ or $B$ for the OB $A$ or $B$, which are respectively associated with angles $\theta_A(\theta_B),\theta'_A(\theta'_B)$, i.e we note $\hat S^{(A)}(\theta_A,\theta'_A)$ and $\hat S^{(B)}(\theta_B,\theta'_B)$. 

The first requirement (noted $R_1$) is that the incoming polarization from space can interfere with the local polarization, and is not fully rejected by the polarized beam splitter. 
For an initial $|H\rangle$ polarization state for the beam on both S/C, the polarization of the beam after leaving the OB $A$ and $B$ is respectively $\hat S^{(A)}(\theta_A,\theta'_A)|H\rangle$ and $\hat S^{(B)}(\theta_B,\theta'_B)|H\rangle$. Then, after travelling in free space, each light beam enters the other S/C and goes through the set of optical elements but with opposite propagation direction. In such a case, light sees the fast axis with opposite angle compared to horizontal plane such that the total polarization of the beam becomes respectively 
\begin{subequations}
\begin{align}\label{eq:LISA_birefringence_cond1_1}
    \hat S^{(B)}(-\theta_B,-\theta'_B)\hat{S}^{(A)}(\theta_A,\theta'_A) &|H\rangle \,\\
    \hat S^{(A)}(-\theta_A,-\theta'_A)\hat{S}^{(B)}(\theta_B,\theta'_B) &|H\rangle \, .
\end{align}
\end{subequations}
$R_1$ implies that both polarizations Eq.~\eqref{eq:LISA_birefringence_cond1_1} are not a pure "horizontal" linear polarization, i.e  
\begin{align}
  R_1 \Rightarrow \left\{ 
\begin{array}{rcl} 
 \hat S^{(B)}(-\theta_B,-\theta'_B)\hat{S}^{(A)}(\theta_A,\theta'_A) |H\rangle &\neq |H\rangle \,\\
    \hat S^{(A)}(-\theta_A,-\theta'_A)\hat{S}^{(B)}(\theta_B,\theta'_B) |H\rangle &\neq |H\rangle \, ,
\end{array} \right.\label{eq:LISA_birefringence_cond1}
\end{align}
up to a constant phase.

The second requirement ($R_2$) for the optical interferometer to work is that the reflected component of light is the same as the initial one, such that it is rejected by the beam splitter, i.e we require that after travelling inside the set $\hat S(\theta,\theta')$, getting reflected on the mirror and travelling back inside $\hat S(\theta,\theta')$ (but with opposite propagation direction, as before), the polarization is "horizontal". This means
\begin{align}\label{eq:LISA_birefringence_cond2}
  R_2 \Rightarrow \left\{ 
\begin{array}{rcl}    
\hat S^{(A)}(-\theta_A,-\theta'_A)\hat{M} S^{(A)}(\theta_A,\theta'_A)  |H\rangle &= |H\rangle \, \\
\hat S^{(B)}(-\theta_B,-\theta'_B)\hat{M} S^{(B)}(\theta_B,\theta'_B)  |H\rangle &= |H\rangle \, ,
\end{array} \right.
\end{align}
still up to a constant phase.

Finally, the additional requirement ($R_3$) to be sensitive to the axion field is that at least one of the two light polarizations in outer space has to be circularly polarized. If we assume the light propagating clockwise is right circularly polarized, then, the light propagating counterclockwise can be anything else, i.e linearly polarized or left polarized state\footnote{If it is linearly polarized or left polarized, the beatnote between the two beams will oscillate at the axion frequency, but the amplitude will be twice as big in the second case.}. Mathematically, 
\begin{align}\label{eq:LISA_birefringence_cond3}
 R_3 \Rightarrow \left\{ 
\begin{array}{rcl}      
    \hat{S}^{(A)}(\theta_A,\theta'_A) |H\rangle &= |R\rangle \,\\
    \hat{S}^{(B)}(\theta_B,\theta'_B) |H\rangle &\neq |R\rangle \, ,
\end{array} \right.
\end{align}
up to a constant phase.

\subsection{Solution}

As mentioned in the main text, a solution to all the requirements of the last section is to replace the last $\hat \lambda_2$ in the OB $A$ by a right polarizer i.e
\begin{subequations}
\begin{align}
    \hat S^{(A)} &= \hat C_R \, \\
    \hat S^{(B)} &= \mathds{1}_2 \, ,
\end{align}
\end{subequations}
where the latter is the identity matrix
We show explicitly how it resolves the various requirements. Starting with $R_3$, we have
\begin{subequations}
\begin{align}
    \hat S^{(A)} |H\rangle &= |R\rangle \, \\
    \hat S^{(B)} |H\rangle &= |H\rangle \neq |R\rangle \, ,
\end{align}
\end{subequations}
and therefore, in free space, one polarization is right circularly polarized and the other linearly polarized such that the axion field impacts only one of the two beams (see Fig.~\ref{fig:opti_bench_LISA_modified}). 

Then, looking at $R_1$, we have
\begin{subequations}
\begin{align}
    \hat S^{(B)}\hat{S}^{(A)} |H\rangle &= |R\rangle \neq |H\rangle \,\\
    \hat S^{(A)}\hat{S}^{(B)} |H\rangle &= |R\rangle \neq |H\rangle  \, ,
\end{align}
\end{subequations}
i.e the light coming from the OB $A$ can interfere in the interferometer in OB $B$ and vice-versa. Indeed, the light leaving OB $A$ is right circularly polarized, arrives in OB $B$ and since a circular polarization is a linear superposition of the two linear polarizations, the vertical component is reflected towards the interferometer, as shown in Fig.~\ref{fig:opti_bench_LISA_modified}. The other beam coming from OB $B$ is in the "horizontal" polarization state in vacuum, arrives on the right polarizer of OB $A$, gets right circularly polarized and similarly as above, its vertical component gets reflected by the polarized beam splitter inside the interferometer (see Fig.~\ref{fig:opti_bench_LISA_modified}).

Finally, let us make sure that no parasitic component of light gets inside the interferometers after getting reflected at the edge of the S/C (requirement $R_2$). First, as we have not modified anything in the OB $B$, we have
 \begin{align}
     \hat S^{(B)}\hat M \hat{S}^{(B)}|H\rangle &= |H\rangle \, ,
 \end{align}
 i.e the reflected component is in the "horizontal" polarization state, and therefore transmitted by the polarized beam splitter.
 Then, it is easy to show that
 \begin{subequations}
 \begin{align}
     \hat C_R \hat M \hat C_R &= 0 \, ,
 \end{align}
such that 
\begin{align}
     \hat S^{(A)}\hat M \hat{S}^{(A)}|H\rangle &= 0 \, ,
 \end{align}
 \end{subequations}
i.e the entire power reflected at the edge of OB $A$ is absorbed by the right polarizer and therefore does not jeopardize the functioning of the interferometer. This can be intuitively understood by the fact that after reflection, the right circularly polarization becomes left circularly polarized and goes back inside the right polarizer, whose role is to extract and transmit only the right polarization of the input. Therefore, nothing from the left polarized parasitic light goes through.

\section{\label{ap:phase_shift}Phase shift induced by vacuum birefringence}

In this appendix, we derive the phase shift induced by vacuum birefringence on the detector arms. 
The axion-photon coupling affects the optical length of the arm via the light travel time $\tau_\rightarrow(t)$ (expressed as a function of the receiving time $t$) from one S/C to the other as
\begin{align}
%L &= \int_0^{\tau_\rightarrow(t)}c(t')dt' =c \tau_\rightarrow(t) \pm  \int_0^{\tau_\rightarrow(t)}\delta c(t')dt' \, ,
    L &= \int^t_{t-\tau_\rightarrow(t)}c(t')dt' =c \tau_\rightarrow(t) \pm  \int^t_{t-\tau_\rightarrow(t)}\delta c(t')dt' \, ,
\end{align}
where $c(t)$ is the time dependent phase velocity of light whose exact form depends on the light polarization Eq.~\eqref{delta_c_axion_photon}, the sign of $\pm$ depends on the right ($+$) or left ($-$) polarization and $\delta c$ is given by Eq.~(\ref{eq:delta_c}). To first order in $\delta c$, the previous equation becomes
\begin{align}
L & =c \tau_\rightarrow(t) \pm  \int^t_{t-L/c}\delta c(t')dt' \, ,
\end{align}
depending on the right or left polarization.

Assuming a right handed photon (i.e. taking the $+$ sign in the previous equation), the photon travel time (as a function of the receiving time) is given by
% \begin{subequations}
\begin{align}\label{time delay}
\tau_\rightarrow(t) = \frac{L}{c} - \int_{t-L/c}^t \frac{\delta c(t')}{c}dt' \, .
\end{align}
% where $\tau_\rightarrow$ is the photon propagation time between two S/C.
% and
% \begin{align}
% \delta c_\rightarrow(t) &=\frac{\sqrt{16\pi G \rho_\mathrm{DM}} E_P g_{a\gamma}}{k c} \sin(\omega_a t+\Phi) \, .
% \end{align} 
% \end{subequations}

We can then interpret this oscillation of phase velocity of light as an oscillation of the distance between the S/C itself, i.e we can write 
\begin{subequations}
\begin{align}\label{eq:tau_ell}
    \tau_\rightarrow(t) &= \frac{L - \delta L(t)}{c} \, ,
\end{align}
where, using Eq.~\eqref{eq:delta_c}
\begin{align}\label{eq:delta-ell}
    &\delta L(t) = \frac{\sqrt{4\pi G \rho_\mathrm{DM}} E_P g_{a\gamma}}{k c} \int_{t-L/c}^t \sin(\omega_at'+\Phi) dt' \\
    &= \frac{\sqrt{16\pi G \rho_\mathrm{DM}} E_P g_{a\gamma}}{\omega_a k c}\sin\left(\frac{\omega_a L}{2c}\right)  \sin\left(\omega_a\left(t-\frac{L}{2c}\right)+\Phi\right) \label{delta_l}
\end{align}
%with 
%\begin{align}\label{eq:f(t)}
%f(t) = \int_{t-\tau_\rightarrow}^t \sin(\omega_at'+\Phi) dt'  \,.
%\end{align}
\end{subequations}
%As we are interested in the first order solution of the phase shift in the perturbation $g_{a\gamma}$, we can solve explicitly $f(t)$ at zeroth order in the perturbation, i.e using $\tau_\rightarrow = \tau_0 \equiv L/c$ in the integral bounds
%\begin{align}
%f(t) &= \frac{2}{\omega_a}\sin\left(\frac{\omega_a L}{2c}\right)\sin\left(\omega_a\left(t-\frac{L}{c}\right)+\Phi\right) \,.
%\end{align}
%such that the variation of the armlength becomes
%\begin{align}
%    \delta L(t) &= \frac{2\sqrt{16\pi G \rho_\mathrm{DM}} E_P g_{a\gamma}}{\omega_a k c}\sin\left(\frac{\omega_a L}{2c}\right)\,\\
%    &\sin\left(\omega_a\left(t-\frac{L}{c}\right)+\Phi\right) \nonumber \, .
%    \label{delta_l}
%\end{align}
Compared to a situation where $\delta L=0$ (typically for linearly polarized light), the phase accumulated by light due to this armlength variation is simply
%\begin{subequations}
\begin{align}
    &\Delta \phi(t) = k\: \delta L(t)\nonumber \,\\
    &=\frac{\sqrt{16\pi G \rho_\mathrm{DM}} E_P g_{a\gamma}}{\omega_a c}\sin\left(\frac{\omega_a L}{2c}\right)\sin\left(\omega_a\left(t-\frac{L}{2c}\right)+\Phi\right) \, .
\end{align}
%\end{subequations}

\bibliographystyle{apsrev4-1}
\bibliography{biblio}

\begin{thebibliography}{49}%
\makeatletter
\providecommand \@ifxundefined [1]{%
 \@ifx{#1\undefined}
}%
\providecommand \@ifnum [1]{%
 \ifnum #1\expandafter \@firstoftwo
 \else \expandafter \@secondoftwo
 \fi
}%
\providecommand \@ifx [1]{%
 \ifx #1\expandafter \@firstoftwo
 \else \expandafter \@secondoftwo
 \fi
}%
\providecommand \natexlab [1]{#1}%
\providecommand \enquote  [1]{``#1''}%
\providecommand \bibnamefont  [1]{#1}%
\providecommand \bibfnamefont [1]{#1}%
\providecommand \citenamefont [1]{#1}%
\providecommand \href@noop [0]{\@secondoftwo}%
\providecommand \href [0]{\begingroup \@sanitize@url \@href}%
\providecommand \@href[1]{\@@startlink{#1}\@@href}%
\providecommand \@@href[1]{\endgroup#1\@@endlink}%
\providecommand \@sanitize@url [0]{\catcode `\\12\catcode `\$12\catcode
  `\&12\catcode `\#12\catcode `\^12\catcode `\_12\catcode `\%12\relax}%
\providecommand \@@startlink[1]{}%
\providecommand \@@endlink[0]{}%
\providecommand \url  [0]{\begingroup\@sanitize@url \@url }%
\providecommand \@url [1]{\endgroup\@href {#1}{\urlprefix }}%
\providecommand \urlprefix  [0]{URL }%
\providecommand \Eprint [0]{\href }%
\providecommand \doibase [0]{http://dx.doi.org/}%
\providecommand \selectlanguage [0]{\@gobble}%
\providecommand \bibinfo  [0]{\@secondoftwo}%
\providecommand \bibfield  [0]{\@secondoftwo}%
\providecommand \translation [1]{[#1]}%
\providecommand \BibitemOpen [0]{}%
\providecommand \bibitemStop [0]{}%
\providecommand \bibitemNoStop [0]{.\EOS\space}%
\providecommand \EOS [0]{\spacefactor3000\relax}%
\providecommand \BibitemShut  [1]{\csname bibitem#1\endcsname}%
\let\auto@bib@innerbib\@empty
%</preamble>
\bibitem [{\citenamefont {Zwicky}(1933)}]{Zwicky33}%
  \BibitemOpen
  \bibfield  {author} {\bibinfo {author} {\bibfnamefont {F.}~\bibnamefont
  {Zwicky}},\ }\href {\doibase 10.1007/s10714-008-0707-4} {\bibfield  {journal}
  {\bibinfo  {journal} {Helv. Phys. Acta}\ }\textbf {\bibinfo {volume} {6}},\
  \bibinfo {pages} {110} (\bibinfo {year} {1933})}\BibitemShut {NoStop}%
\bibitem [{\citenamefont {Bertone}\ and\ \citenamefont
  {Tait}(2018)}]{Bertone18}%
  \BibitemOpen
  \bibfield  {author} {\bibinfo {author} {\bibfnamefont {G.}~\bibnamefont
  {Bertone}}\ and\ \bibinfo {author} {\bibfnamefont {T.~M.~P.}\ \bibnamefont
  {Tait}},\ }\href {\doibase 10.1038/s41586-018-0542-z} {\bibfield  {journal}
  {\bibinfo  {journal} {Nature}\ }\textbf {\bibinfo {volume} {562}},\ \bibinfo
  {pages} {51–56} (\bibinfo {year} {2018})}\BibitemShut {NoStop}%
\bibitem [{\citenamefont {Battaglieri}\ \emph {et~al.}(2017)\citenamefont
  {Battaglieri}, \citenamefont {Belloni}, \citenamefont {Chou}, \citenamefont
  {Cushman}, \citenamefont {Echenard}, \citenamefont {Essig}, \citenamefont
  {Estrada}, \citenamefont {Feng}, \citenamefont {Flaugher}, \citenamefont
  {Fox}, \citenamefont {Graham}, \citenamefont {Hall}, \citenamefont {Harnik},
  \citenamefont {Hewett}, \citenamefont {Incandela}, \citenamefont {Izaguirre},
  \citenamefont {McKinsey}, \citenamefont {Pyle}, \citenamefont {Roe},
  \citenamefont {Rybka}, \citenamefont {Sikivie}, \citenamefont {Tait},
  \citenamefont {Toro}, \citenamefont {Water}, \citenamefont {Weiner},
  \citenamefont {Zurek}, \citenamefont {Adelberger}, \citenamefont {Afanasev},
  \citenamefont {Alexander}, \citenamefont {Alexander}, \citenamefont
  {Antochi}, \citenamefont {Asner}, \citenamefont {Baer}, \citenamefont
  {Banerjee}, \citenamefont {Baracchini}, \citenamefont {Barbeau},
  \citenamefont {Barrow}, \citenamefont {Bastidon}, \citenamefont {Battat},
  \citenamefont {Benson}, \citenamefont {Berlin}, \citenamefont {Bird},
  \citenamefont {Blinov}, \citenamefont {Boddy}, \citenamefont {Bondi},
  \citenamefont {Bonivento}, \citenamefont {Boulay}, \citenamefont {Boyce},
  \citenamefont {Brodeur}, \citenamefont {Broussard}, \citenamefont {Budnik},
  \citenamefont {Bunting}, \citenamefont {Caffee}, \citenamefont {Caiazza},
  \citenamefont {Campbell}, \citenamefont {Cao}, \citenamefont {Carosi},
  \citenamefont {Carpinelli}, \citenamefont {Cavoto}, \citenamefont
  {Celentano}, \citenamefont {Chang}, \citenamefont {Chattopadhyay},
  \citenamefont {Chavarria}, \citenamefont {Chen}, \citenamefont {Clark},
  \citenamefont {Clarke}, \citenamefont {Colegrove}, \citenamefont {Coleman},
  \citenamefont {Cooke}, \citenamefont {Cooper}, \citenamefont {Crisler},
  \citenamefont {Crivelli}, \citenamefont {D'Eramo}, \citenamefont {D'Urso},
  \citenamefont {Dahl}, \citenamefont {Dawson}, \citenamefont {Napoli},
  \citenamefont {Vita}, \citenamefont {DeNiverville}, \citenamefont {Derenzo},
  \citenamefont {Crescenzo}, \citenamefont {Marco}, \citenamefont {Dienes},
  \citenamefont {Diwan}, \citenamefont {Dongwi}, \citenamefont {Drlica-Wagner},
  \citenamefont {Ellis}, \citenamefont {Ezeribe}, \citenamefont {Farrar},
  \citenamefont {Ferrer}, \citenamefont {Figueroa-Feliciano}, \citenamefont
  {Filippi}, \citenamefont {Fiorillo}, \citenamefont {Fornal}, \citenamefont
  {Freyberger}, \citenamefont {Frugiuele}, \citenamefont {Galbiati},
  \citenamefont {Galon}, \citenamefont {Gardner}, \citenamefont {Geraci},
  \citenamefont {Gerbier}, \citenamefont {Graham}, \citenamefont
  {Gschwendtner}, \citenamefont {Hearty}, \citenamefont {Heise}, \citenamefont
  {Henning}, \citenamefont {Hill}, \citenamefont {Hitlin}, \citenamefont
  {Hochberg}, \citenamefont {Hogan}, \citenamefont {Holtrop}, \citenamefont
  {Hong}, \citenamefont {Hossbach}, \citenamefont {Humensky}, \citenamefont
  {Ilten}, \citenamefont {Irwin}, \citenamefont {Jaros}, \citenamefont
  {Johnson}, \citenamefont {Jones}, \citenamefont {Kahn}, \citenamefont
  {Kalantarians}, \citenamefont {Kaplinghat}, \citenamefont {Khatiwada},
  \citenamefont {Knapen}, \citenamefont {Kohl}, \citenamefont {Kouvaris},
  \citenamefont {Kozaczuk}, \citenamefont {Krnjaic}, \citenamefont
  {Kubarovsky}, \citenamefont {Kuflik}, \citenamefont {Kusenko}, \citenamefont
  {Lang}, \citenamefont {Leach}, \citenamefont {Lin}, \citenamefont {Lisanti},
  \citenamefont {Liu}, \citenamefont {Liu}, \citenamefont {Liu}, \citenamefont
  {Loomba}, \citenamefont {Lykken}, \citenamefont {Mack}, \citenamefont {Mans},
  \citenamefont {Maris}, \citenamefont {Markiewicz}, \citenamefont {Marsicano},
  \citenamefont {Martoff}, \citenamefont {Mazzitelli}, \citenamefont {McCabe},
  \citenamefont {McDermott}, \citenamefont {McDonald}, \citenamefont
  {McKinnon}, \citenamefont {Mei}, \citenamefont {Melia}, \citenamefont
  {Miller}, \citenamefont {Miuchi}, \citenamefont {Nazeer}, \citenamefont
  {Moreno}, \citenamefont {Morozov}, \citenamefont {Mouton}, \citenamefont
  {Mueller}, \citenamefont {Murphy}, \citenamefont {Neilson}, \citenamefont
  {Nelson}, \citenamefont {Neu}, \citenamefont {Nosochkov}, \citenamefont
  {O'Hare}, \citenamefont {Oblath}, \citenamefont {Orrell}, \citenamefont
  {Ouellet}, \citenamefont {Pastore}, \citenamefont {Paul}, \citenamefont
  {Perelstein}, \citenamefont {Peter}, \citenamefont {Phan}, \citenamefont
  {Phinney}, \citenamefont {Pivovaroff}, \citenamefont {Pocar}, \citenamefont
  {Pospelov}, \citenamefont {Pradler}, \citenamefont {Privitera}, \citenamefont
  {Profumo}, \citenamefont {Raggi}, \citenamefont {Rajendran}, \citenamefont
  {Randazzo}, \citenamefont {Raubenheimer}, \citenamefont {Regenfus},
  \citenamefont {Renshaw}, \citenamefont {Ritz}, \citenamefont {Rizzo},
  \citenamefont {Rosenberg}, \citenamefont {Rubbia}, \citenamefont {Rybolt},
  \citenamefont {Saab}, \citenamefont {Safdi}, \citenamefont {Santopinto},
  \citenamefont {Scarff}, \citenamefont {Schneider}, \citenamefont {Schuster},
  \citenamefont {Seidel}, \citenamefont {Sekiya}, \citenamefont {Seong},
  \citenamefont {Simi}, \citenamefont {Sipala}, \citenamefont {Slatyer},
  \citenamefont {Slone}, \citenamefont {Smith}, \citenamefont {Smolinsky},
  \citenamefont {Snowden-Ifft}, \citenamefont {Solt}, \citenamefont
  {Sonnenschein}, \citenamefont {Sorensen}, \citenamefont {Spooner},
  \citenamefont {Srivastava}, \citenamefont {Stancu}, \citenamefont {Strigari},
  \citenamefont {Strube}, \citenamefont {Sushkov}, \citenamefont {Szydagis},
  \citenamefont {Tanedo}, \citenamefont {Tanner}, \citenamefont {Tayloe},
  \citenamefont {Terrano}, \citenamefont {Thaler}, \citenamefont {Thomas},
  \citenamefont {Thorpe}, \citenamefont {Thorpe}, \citenamefont {Tiffenberg},
  \citenamefont {Tran}, \citenamefont {Trovato}, \citenamefont {Tully},
  \citenamefont {Tyson}, \citenamefont {Vachaspati}, \citenamefont {Vahsen},
  \citenamefont {van Bibber}, \citenamefont {Vandenbroucke}, \citenamefont
  {Villano}, \citenamefont {Volansky}, \citenamefont {Wang}, \citenamefont
  {Ward}, \citenamefont {Wester}, \citenamefont {Whitbeck}, \citenamefont
  {Williams}, \citenamefont {Wing}, \citenamefont {Winslow}, \citenamefont
  {Wojtsekhowski}, \citenamefont {Yu}, \citenamefont {Yu}, \citenamefont {Yu},
  \citenamefont {Zhang}, \citenamefont {Zhao},\ and\ \citenamefont
  {Zhong}}]{Battaglieri17}%
  \BibitemOpen
  \bibfield  {author} {\bibinfo {author} {\bibfnamefont {M.}~\bibnamefont
  {Battaglieri}}, \bibinfo {author} {\bibfnamefont {A.}~\bibnamefont
  {Belloni}}, \bibinfo {author} {\bibfnamefont {A.}~\bibnamefont {Chou}},
  \bibinfo {author} {\bibfnamefont {P.}~\bibnamefont {Cushman}}, \bibinfo
  {author} {\bibfnamefont {B.}~\bibnamefont {Echenard}}, \bibinfo {author}
  {\bibfnamefont {R.}~\bibnamefont {Essig}}, \bibinfo {author} {\bibfnamefont
  {J.}~\bibnamefont {Estrada}}, \bibinfo {author} {\bibfnamefont {J.~L.}\
  \bibnamefont {Feng}}, \bibinfo {author} {\bibfnamefont {B.}~\bibnamefont
  {Flaugher}}, \bibinfo {author} {\bibfnamefont {P.~J.}\ \bibnamefont {Fox}},
  \bibinfo {author} {\bibfnamefont {P.}~\bibnamefont {Graham}}, \bibinfo
  {author} {\bibfnamefont {C.}~\bibnamefont {Hall}}, \bibinfo {author}
  {\bibfnamefont {R.}~\bibnamefont {Harnik}}, \bibinfo {author} {\bibfnamefont
  {J.}~\bibnamefont {Hewett}}, \bibinfo {author} {\bibfnamefont
  {J.}~\bibnamefont {Incandela}}, \bibinfo {author} {\bibfnamefont
  {E.}~\bibnamefont {Izaguirre}}, \bibinfo {author} {\bibfnamefont
  {D.}~\bibnamefont {McKinsey}}, \bibinfo {author} {\bibfnamefont
  {M.}~\bibnamefont {Pyle}}, \bibinfo {author} {\bibfnamefont {N.}~\bibnamefont
  {Roe}}, \bibinfo {author} {\bibfnamefont {G.}~\bibnamefont {Rybka}}, \bibinfo
  {author} {\bibfnamefont {P.}~\bibnamefont {Sikivie}}, \bibinfo {author}
  {\bibfnamefont {T.~M.~P.}\ \bibnamefont {Tait}}, \bibinfo {author}
  {\bibfnamefont {N.}~\bibnamefont {Toro}}, \bibinfo {author} {\bibfnamefont
  {R.~V.~D.}\ \bibnamefont {Water}}, \bibinfo {author} {\bibfnamefont
  {N.}~\bibnamefont {Weiner}}, \bibinfo {author} {\bibfnamefont
  {K.}~\bibnamefont {Zurek}}, \bibinfo {author} {\bibfnamefont
  {E.}~\bibnamefont {Adelberger}}, \bibinfo {author} {\bibfnamefont
  {A.}~\bibnamefont {Afanasev}}, \bibinfo {author} {\bibfnamefont
  {D.}~\bibnamefont {Alexander}}, \bibinfo {author} {\bibfnamefont
  {J.}~\bibnamefont {Alexander}}, \bibinfo {author} {\bibfnamefont {V.~C.}\
  \bibnamefont {Antochi}}, \bibinfo {author} {\bibfnamefont {D.~M.}\
  \bibnamefont {Asner}}, \bibinfo {author} {\bibfnamefont {H.}~\bibnamefont
  {Baer}}, \bibinfo {author} {\bibfnamefont {D.}~\bibnamefont {Banerjee}},
  \bibinfo {author} {\bibfnamefont {E.}~\bibnamefont {Baracchini}}, \bibinfo
  {author} {\bibfnamefont {P.}~\bibnamefont {Barbeau}}, \bibinfo {author}
  {\bibfnamefont {J.}~\bibnamefont {Barrow}}, \bibinfo {author} {\bibfnamefont
  {N.}~\bibnamefont {Bastidon}}, \bibinfo {author} {\bibfnamefont
  {J.}~\bibnamefont {Battat}}, \bibinfo {author} {\bibfnamefont
  {S.}~\bibnamefont {Benson}}, \bibinfo {author} {\bibfnamefont
  {A.}~\bibnamefont {Berlin}}, \bibinfo {author} {\bibfnamefont
  {M.}~\bibnamefont {Bird}}, \bibinfo {author} {\bibfnamefont {N.}~\bibnamefont
  {Blinov}}, \bibinfo {author} {\bibfnamefont {K.~K.}\ \bibnamefont {Boddy}},
  \bibinfo {author} {\bibfnamefont {M.}~\bibnamefont {Bondi}}, \bibinfo
  {author} {\bibfnamefont {W.~M.}\ \bibnamefont {Bonivento}}, \bibinfo {author}
  {\bibfnamefont {M.}~\bibnamefont {Boulay}}, \bibinfo {author} {\bibfnamefont
  {J.}~\bibnamefont {Boyce}}, \bibinfo {author} {\bibfnamefont
  {M.}~\bibnamefont {Brodeur}}, \bibinfo {author} {\bibfnamefont
  {L.}~\bibnamefont {Broussard}}, \bibinfo {author} {\bibfnamefont
  {R.}~\bibnamefont {Budnik}}, \bibinfo {author} {\bibfnamefont
  {P.}~\bibnamefont {Bunting}}, \bibinfo {author} {\bibfnamefont
  {M.}~\bibnamefont {Caffee}}, \bibinfo {author} {\bibfnamefont {S.~S.}\
  \bibnamefont {Caiazza}}, \bibinfo {author} {\bibfnamefont {S.}~\bibnamefont
  {Campbell}}, \bibinfo {author} {\bibfnamefont {T.}~\bibnamefont {Cao}},
  \bibinfo {author} {\bibfnamefont {G.}~\bibnamefont {Carosi}}, \bibinfo
  {author} {\bibfnamefont {M.}~\bibnamefont {Carpinelli}}, \bibinfo {author}
  {\bibfnamefont {G.}~\bibnamefont {Cavoto}}, \bibinfo {author} {\bibfnamefont
  {A.}~\bibnamefont {Celentano}}, \bibinfo {author} {\bibfnamefont {J.~H.}\
  \bibnamefont {Chang}}, \bibinfo {author} {\bibfnamefont {S.}~\bibnamefont
  {Chattopadhyay}}, \bibinfo {author} {\bibfnamefont {A.}~\bibnamefont
  {Chavarria}}, \bibinfo {author} {\bibfnamefont {C.-Y.}\ \bibnamefont {Chen}},
  \bibinfo {author} {\bibfnamefont {K.}~\bibnamefont {Clark}}, \bibinfo
  {author} {\bibfnamefont {J.}~\bibnamefont {Clarke}}, \bibinfo {author}
  {\bibfnamefont {O.}~\bibnamefont {Colegrove}}, \bibinfo {author}
  {\bibfnamefont {J.}~\bibnamefont {Coleman}}, \bibinfo {author} {\bibfnamefont
  {D.}~\bibnamefont {Cooke}}, \bibinfo {author} {\bibfnamefont
  {R.}~\bibnamefont {Cooper}}, \bibinfo {author} {\bibfnamefont
  {M.}~\bibnamefont {Crisler}}, \bibinfo {author} {\bibfnamefont
  {P.}~\bibnamefont {Crivelli}}, \bibinfo {author} {\bibfnamefont
  {F.}~\bibnamefont {D'Eramo}}, \bibinfo {author} {\bibfnamefont
  {D.}~\bibnamefont {D'Urso}}, \bibinfo {author} {\bibfnamefont
  {E.}~\bibnamefont {Dahl}}, \bibinfo {author} {\bibfnamefont {W.}~\bibnamefont
  {Dawson}}, \bibinfo {author} {\bibfnamefont {M.~D.}\ \bibnamefont {Napoli}},
  \bibinfo {author} {\bibfnamefont {R.~D.}\ \bibnamefont {Vita}}, \bibinfo
  {author} {\bibfnamefont {P.}~\bibnamefont {DeNiverville}}, \bibinfo {author}
  {\bibfnamefont {S.}~\bibnamefont {Derenzo}}, \bibinfo {author} {\bibfnamefont
  {A.~D.}\ \bibnamefont {Crescenzo}}, \bibinfo {author} {\bibfnamefont {E.~D.}\
  \bibnamefont {Marco}}, \bibinfo {author} {\bibfnamefont {K.~R.}\ \bibnamefont
  {Dienes}}, \bibinfo {author} {\bibfnamefont {M.}~\bibnamefont {Diwan}},
  \bibinfo {author} {\bibfnamefont {D.~H.}\ \bibnamefont {Dongwi}}, \bibinfo
  {author} {\bibfnamefont {A.}~\bibnamefont {Drlica-Wagner}}, \bibinfo {author}
  {\bibfnamefont {S.}~\bibnamefont {Ellis}}, \bibinfo {author} {\bibfnamefont
  {A.~C.}\ \bibnamefont {Ezeribe}}, \bibinfo {author} {\bibfnamefont
  {G.}~\bibnamefont {Farrar}}, \bibinfo {author} {\bibfnamefont
  {F.}~\bibnamefont {Ferrer}}, \bibinfo {author} {\bibfnamefont
  {E.}~\bibnamefont {Figueroa-Feliciano}}, \bibinfo {author} {\bibfnamefont
  {A.}~\bibnamefont {Filippi}}, \bibinfo {author} {\bibfnamefont
  {G.}~\bibnamefont {Fiorillo}}, \bibinfo {author} {\bibfnamefont
  {B.}~\bibnamefont {Fornal}}, \bibinfo {author} {\bibfnamefont
  {A.}~\bibnamefont {Freyberger}}, \bibinfo {author} {\bibfnamefont
  {C.}~\bibnamefont {Frugiuele}}, \bibinfo {author} {\bibfnamefont
  {C.}~\bibnamefont {Galbiati}}, \bibinfo {author} {\bibfnamefont
  {I.}~\bibnamefont {Galon}}, \bibinfo {author} {\bibfnamefont
  {S.}~\bibnamefont {Gardner}}, \bibinfo {author} {\bibfnamefont
  {A.}~\bibnamefont {Geraci}}, \bibinfo {author} {\bibfnamefont
  {G.}~\bibnamefont {Gerbier}}, \bibinfo {author} {\bibfnamefont
  {M.}~\bibnamefont {Graham}}, \bibinfo {author} {\bibfnamefont
  {E.}~\bibnamefont {Gschwendtner}}, \bibinfo {author} {\bibfnamefont
  {C.}~\bibnamefont {Hearty}}, \bibinfo {author} {\bibfnamefont
  {J.}~\bibnamefont {Heise}}, \bibinfo {author} {\bibfnamefont
  {R.}~\bibnamefont {Henning}}, \bibinfo {author} {\bibfnamefont {R.~J.}\
  \bibnamefont {Hill}}, \bibinfo {author} {\bibfnamefont {D.}~\bibnamefont
  {Hitlin}}, \bibinfo {author} {\bibfnamefont {Y.}~\bibnamefont {Hochberg}},
  \bibinfo {author} {\bibfnamefont {J.}~\bibnamefont {Hogan}}, \bibinfo
  {author} {\bibfnamefont {M.}~\bibnamefont {Holtrop}}, \bibinfo {author}
  {\bibfnamefont {Z.}~\bibnamefont {Hong}}, \bibinfo {author} {\bibfnamefont
  {T.}~\bibnamefont {Hossbach}}, \bibinfo {author} {\bibfnamefont {T.~B.}\
  \bibnamefont {Humensky}}, \bibinfo {author} {\bibfnamefont {P.}~\bibnamefont
  {Ilten}}, \bibinfo {author} {\bibfnamefont {K.}~\bibnamefont {Irwin}},
  \bibinfo {author} {\bibfnamefont {J.}~\bibnamefont {Jaros}}, \bibinfo
  {author} {\bibfnamefont {R.}~\bibnamefont {Johnson}}, \bibinfo {author}
  {\bibfnamefont {M.}~\bibnamefont {Jones}}, \bibinfo {author} {\bibfnamefont
  {Y.}~\bibnamefont {Kahn}}, \bibinfo {author} {\bibfnamefont {N.}~\bibnamefont
  {Kalantarians}}, \bibinfo {author} {\bibfnamefont {M.}~\bibnamefont
  {Kaplinghat}}, \bibinfo {author} {\bibfnamefont {R.}~\bibnamefont
  {Khatiwada}}, \bibinfo {author} {\bibfnamefont {S.}~\bibnamefont {Knapen}},
  \bibinfo {author} {\bibfnamefont {M.}~\bibnamefont {Kohl}}, \bibinfo {author}
  {\bibfnamefont {C.}~\bibnamefont {Kouvaris}}, \bibinfo {author}
  {\bibfnamefont {J.}~\bibnamefont {Kozaczuk}}, \bibinfo {author}
  {\bibfnamefont {G.}~\bibnamefont {Krnjaic}}, \bibinfo {author} {\bibfnamefont
  {V.}~\bibnamefont {Kubarovsky}}, \bibinfo {author} {\bibfnamefont
  {E.}~\bibnamefont {Kuflik}}, \bibinfo {author} {\bibfnamefont
  {A.}~\bibnamefont {Kusenko}}, \bibinfo {author} {\bibfnamefont
  {R.}~\bibnamefont {Lang}}, \bibinfo {author} {\bibfnamefont {K.}~\bibnamefont
  {Leach}}, \bibinfo {author} {\bibfnamefont {T.}~\bibnamefont {Lin}}, \bibinfo
  {author} {\bibfnamefont {M.}~\bibnamefont {Lisanti}}, \bibinfo {author}
  {\bibfnamefont {J.}~\bibnamefont {Liu}}, \bibinfo {author} {\bibfnamefont
  {K.}~\bibnamefont {Liu}}, \bibinfo {author} {\bibfnamefont {M.}~\bibnamefont
  {Liu}}, \bibinfo {author} {\bibfnamefont {D.}~\bibnamefont {Loomba}},
  \bibinfo {author} {\bibfnamefont {J.}~\bibnamefont {Lykken}}, \bibinfo
  {author} {\bibfnamefont {K.}~\bibnamefont {Mack}}, \bibinfo {author}
  {\bibfnamefont {J.}~\bibnamefont {Mans}}, \bibinfo {author} {\bibfnamefont
  {H.}~\bibnamefont {Maris}}, \bibinfo {author} {\bibfnamefont
  {T.}~\bibnamefont {Markiewicz}}, \bibinfo {author} {\bibfnamefont
  {L.}~\bibnamefont {Marsicano}}, \bibinfo {author} {\bibfnamefont {C.~J.}\
  \bibnamefont {Martoff}}, \bibinfo {author} {\bibfnamefont {G.}~\bibnamefont
  {Mazzitelli}}, \bibinfo {author} {\bibfnamefont {C.}~\bibnamefont {McCabe}},
  \bibinfo {author} {\bibfnamefont {S.~D.}\ \bibnamefont {McDermott}}, \bibinfo
  {author} {\bibfnamefont {A.}~\bibnamefont {McDonald}}, \bibinfo {author}
  {\bibfnamefont {B.}~\bibnamefont {McKinnon}}, \bibinfo {author}
  {\bibfnamefont {D.}~\bibnamefont {Mei}}, \bibinfo {author} {\bibfnamefont
  {T.}~\bibnamefont {Melia}}, \bibinfo {author} {\bibfnamefont {G.~A.}\
  \bibnamefont {Miller}}, \bibinfo {author} {\bibfnamefont {K.}~\bibnamefont
  {Miuchi}}, \bibinfo {author} {\bibfnamefont {S.~M.~P.}\ \bibnamefont
  {Nazeer}}, \bibinfo {author} {\bibfnamefont {O.}~\bibnamefont {Moreno}},
  \bibinfo {author} {\bibfnamefont {V.}~\bibnamefont {Morozov}}, \bibinfo
  {author} {\bibfnamefont {F.}~\bibnamefont {Mouton}}, \bibinfo {author}
  {\bibfnamefont {H.}~\bibnamefont {Mueller}}, \bibinfo {author} {\bibfnamefont
  {A.}~\bibnamefont {Murphy}}, \bibinfo {author} {\bibfnamefont
  {R.}~\bibnamefont {Neilson}}, \bibinfo {author} {\bibfnamefont
  {T.}~\bibnamefont {Nelson}}, \bibinfo {author} {\bibfnamefont
  {C.}~\bibnamefont {Neu}}, \bibinfo {author} {\bibfnamefont {Y.}~\bibnamefont
  {Nosochkov}}, \bibinfo {author} {\bibfnamefont {C.}~\bibnamefont {O'Hare}},
  \bibinfo {author} {\bibfnamefont {N.}~\bibnamefont {Oblath}}, \bibinfo
  {author} {\bibfnamefont {J.}~\bibnamefont {Orrell}}, \bibinfo {author}
  {\bibfnamefont {J.}~\bibnamefont {Ouellet}}, \bibinfo {author} {\bibfnamefont
  {S.}~\bibnamefont {Pastore}}, \bibinfo {author} {\bibfnamefont
  {S.}~\bibnamefont {Paul}}, \bibinfo {author} {\bibfnamefont {M.}~\bibnamefont
  {Perelstein}}, \bibinfo {author} {\bibfnamefont {A.}~\bibnamefont {Peter}},
  \bibinfo {author} {\bibfnamefont {N.}~\bibnamefont {Phan}}, \bibinfo {author}
  {\bibfnamefont {N.}~\bibnamefont {Phinney}}, \bibinfo {author} {\bibfnamefont
  {M.}~\bibnamefont {Pivovaroff}}, \bibinfo {author} {\bibfnamefont
  {A.}~\bibnamefont {Pocar}}, \bibinfo {author} {\bibfnamefont
  {M.}~\bibnamefont {Pospelov}}, \bibinfo {author} {\bibfnamefont
  {J.}~\bibnamefont {Pradler}}, \bibinfo {author} {\bibfnamefont
  {P.}~\bibnamefont {Privitera}}, \bibinfo {author} {\bibfnamefont
  {S.}~\bibnamefont {Profumo}}, \bibinfo {author} {\bibfnamefont
  {M.}~\bibnamefont {Raggi}}, \bibinfo {author} {\bibfnamefont
  {S.}~\bibnamefont {Rajendran}}, \bibinfo {author} {\bibfnamefont
  {N.}~\bibnamefont {Randazzo}}, \bibinfo {author} {\bibfnamefont
  {T.}~\bibnamefont {Raubenheimer}}, \bibinfo {author} {\bibfnamefont
  {C.}~\bibnamefont {Regenfus}}, \bibinfo {author} {\bibfnamefont
  {A.}~\bibnamefont {Renshaw}}, \bibinfo {author} {\bibfnamefont
  {A.}~\bibnamefont {Ritz}}, \bibinfo {author} {\bibfnamefont {T.}~\bibnamefont
  {Rizzo}}, \bibinfo {author} {\bibfnamefont {L.}~\bibnamefont {Rosenberg}},
  \bibinfo {author} {\bibfnamefont {A.}~\bibnamefont {Rubbia}}, \bibinfo
  {author} {\bibfnamefont {B.}~\bibnamefont {Rybolt}}, \bibinfo {author}
  {\bibfnamefont {T.}~\bibnamefont {Saab}}, \bibinfo {author} {\bibfnamefont
  {B.~R.}\ \bibnamefont {Safdi}}, \bibinfo {author} {\bibfnamefont
  {E.}~\bibnamefont {Santopinto}}, \bibinfo {author} {\bibfnamefont
  {A.}~\bibnamefont {Scarff}}, \bibinfo {author} {\bibfnamefont
  {M.}~\bibnamefont {Schneider}}, \bibinfo {author} {\bibfnamefont
  {P.}~\bibnamefont {Schuster}}, \bibinfo {author} {\bibfnamefont
  {G.}~\bibnamefont {Seidel}}, \bibinfo {author} {\bibfnamefont
  {H.}~\bibnamefont {Sekiya}}, \bibinfo {author} {\bibfnamefont
  {I.}~\bibnamefont {Seong}}, \bibinfo {author} {\bibfnamefont
  {G.}~\bibnamefont {Simi}}, \bibinfo {author} {\bibfnamefont {V.}~\bibnamefont
  {Sipala}}, \bibinfo {author} {\bibfnamefont {T.}~\bibnamefont {Slatyer}},
  \bibinfo {author} {\bibfnamefont {O.}~\bibnamefont {Slone}}, \bibinfo
  {author} {\bibfnamefont {P.~F.}\ \bibnamefont {Smith}}, \bibinfo {author}
  {\bibfnamefont {J.}~\bibnamefont {Smolinsky}}, \bibinfo {author}
  {\bibfnamefont {D.}~\bibnamefont {Snowden-Ifft}}, \bibinfo {author}
  {\bibfnamefont {M.}~\bibnamefont {Solt}}, \bibinfo {author} {\bibfnamefont
  {A.}~\bibnamefont {Sonnenschein}}, \bibinfo {author} {\bibfnamefont
  {P.}~\bibnamefont {Sorensen}}, \bibinfo {author} {\bibfnamefont
  {N.}~\bibnamefont {Spooner}}, \bibinfo {author} {\bibfnamefont
  {B.}~\bibnamefont {Srivastava}}, \bibinfo {author} {\bibfnamefont
  {I.}~\bibnamefont {Stancu}}, \bibinfo {author} {\bibfnamefont
  {L.}~\bibnamefont {Strigari}}, \bibinfo {author} {\bibfnamefont
  {J.}~\bibnamefont {Strube}}, \bibinfo {author} {\bibfnamefont {A.~O.}\
  \bibnamefont {Sushkov}}, \bibinfo {author} {\bibfnamefont {M.}~\bibnamefont
  {Szydagis}}, \bibinfo {author} {\bibfnamefont {P.}~\bibnamefont {Tanedo}},
  \bibinfo {author} {\bibfnamefont {D.}~\bibnamefont {Tanner}}, \bibinfo
  {author} {\bibfnamefont {R.}~\bibnamefont {Tayloe}}, \bibinfo {author}
  {\bibfnamefont {W.}~\bibnamefont {Terrano}}, \bibinfo {author} {\bibfnamefont
  {J.}~\bibnamefont {Thaler}}, \bibinfo {author} {\bibfnamefont
  {B.}~\bibnamefont {Thomas}}, \bibinfo {author} {\bibfnamefont
  {B.}~\bibnamefont {Thorpe}}, \bibinfo {author} {\bibfnamefont
  {T.}~\bibnamefont {Thorpe}}, \bibinfo {author} {\bibfnamefont
  {J.}~\bibnamefont {Tiffenberg}}, \bibinfo {author} {\bibfnamefont
  {N.}~\bibnamefont {Tran}}, \bibinfo {author} {\bibfnamefont {M.}~\bibnamefont
  {Trovato}}, \bibinfo {author} {\bibfnamefont {C.}~\bibnamefont {Tully}},
  \bibinfo {author} {\bibfnamefont {T.}~\bibnamefont {Tyson}}, \bibinfo
  {author} {\bibfnamefont {T.}~\bibnamefont {Vachaspati}}, \bibinfo {author}
  {\bibfnamefont {S.}~\bibnamefont {Vahsen}}, \bibinfo {author} {\bibfnamefont
  {K.}~\bibnamefont {van Bibber}}, \bibinfo {author} {\bibfnamefont
  {J.}~\bibnamefont {Vandenbroucke}}, \bibinfo {author} {\bibfnamefont
  {A.}~\bibnamefont {Villano}}, \bibinfo {author} {\bibfnamefont
  {T.}~\bibnamefont {Volansky}}, \bibinfo {author} {\bibfnamefont
  {G.}~\bibnamefont {Wang}}, \bibinfo {author} {\bibfnamefont {T.}~\bibnamefont
  {Ward}}, \bibinfo {author} {\bibfnamefont {W.}~\bibnamefont {Wester}},
  \bibinfo {author} {\bibfnamefont {A.}~\bibnamefont {Whitbeck}}, \bibinfo
  {author} {\bibfnamefont {D.~A.}\ \bibnamefont {Williams}}, \bibinfo {author}
  {\bibfnamefont {M.}~\bibnamefont {Wing}}, \bibinfo {author} {\bibfnamefont
  {L.}~\bibnamefont {Winslow}}, \bibinfo {author} {\bibfnamefont
  {B.}~\bibnamefont {Wojtsekhowski}}, \bibinfo {author} {\bibfnamefont {H.-B.}\
  \bibnamefont {Yu}}, \bibinfo {author} {\bibfnamefont {S.-S.}\ \bibnamefont
  {Yu}}, \bibinfo {author} {\bibfnamefont {T.-T.}\ \bibnamefont {Yu}}, \bibinfo
  {author} {\bibfnamefont {X.}~\bibnamefont {Zhang}}, \bibinfo {author}
  {\bibfnamefont {Y.}~\bibnamefont {Zhao}}, \ and\ \bibinfo {author}
  {\bibfnamefont {Y.-M.}\ \bibnamefont {Zhong}},\ }\href@noop {} {\enquote
  {\bibinfo {title} {Us cosmic visions: New ideas in dark matter 2017:
  Community report},}\ } (\bibinfo {year} {2017}),\ \Eprint
  {http://arxiv.org/abs/1707.04591} {arXiv:1707.04591 [hep-ph]} \BibitemShut
  {NoStop}%
\bibitem [{\citenamefont {Carr}\ \emph {et~al.}(2016)\citenamefont {Carr},
  \citenamefont {Kühnel},\ and\ \citenamefont {Sandstad}}]{Carr16}%
  \BibitemOpen
  \bibfield  {author} {\bibinfo {author} {\bibfnamefont {B.}~\bibnamefont
  {Carr}}, \bibinfo {author} {\bibfnamefont {F.}~\bibnamefont {Kühnel}}, \
  and\ \bibinfo {author} {\bibfnamefont {M.}~\bibnamefont {Sandstad}},\ }\href
  {\doibase 10.1103/physrevd.94.083504} {\bibfield  {journal} {\bibinfo
  {journal} {Physical Review D}\ }\textbf {\bibinfo {volume} {94}} (\bibinfo
  {year} {2016}),\ 10.1103/physrevd.94.083504}\BibitemShut {NoStop}%
\bibitem [{\citenamefont {Safronova}\ \emph {et~al.}(2018)\citenamefont
  {Safronova}, \citenamefont {Budker}, \citenamefont {DeMille}, \citenamefont
  {Kimball}, \citenamefont {Derevianko},\ and\ \citenamefont
  {Clark}}]{Safronova18}%
  \BibitemOpen
  \bibfield  {author} {\bibinfo {author} {\bibfnamefont {M.~S.}\ \bibnamefont
  {Safronova}}, \bibinfo {author} {\bibfnamefont {D.}~\bibnamefont {Budker}},
  \bibinfo {author} {\bibfnamefont {D.}~\bibnamefont {DeMille}}, \bibinfo
  {author} {\bibfnamefont {D.~F.~J.}\ \bibnamefont {Kimball}}, \bibinfo
  {author} {\bibfnamefont {A.}~\bibnamefont {Derevianko}}, \ and\ \bibinfo
  {author} {\bibfnamefont {C.~W.}\ \bibnamefont {Clark}},\ }\href {\doibase
  10.1103/RevModPhys.90.025008} {\bibfield  {journal} {\bibinfo  {journal}
  {Rev. Mod. Phys.}\ }\textbf {\bibinfo {volume} {90}},\ \bibinfo {pages}
  {025008} (\bibinfo {year} {2018})}\BibitemShut {NoStop}%
\bibitem [{\citenamefont {Weinberg}(1978)}]{Weinberg78}%
  \BibitemOpen
  \bibfield  {author} {\bibinfo {author} {\bibfnamefont {S.}~\bibnamefont
  {Weinberg}},\ }\href {\doibase 10.1103/PhysRevLett.40.223} {\bibfield
  {journal} {\bibinfo  {journal} {Phys. Rev. Lett.}\ }\textbf {\bibinfo
  {volume} {40}},\ \bibinfo {pages} {223} (\bibinfo {year} {1978})}\BibitemShut
  {NoStop}%
\bibitem [{\citenamefont {Wilczek}(1978)}]{Wilczek78}%
  \BibitemOpen
  \bibfield  {author} {\bibinfo {author} {\bibfnamefont {F.}~\bibnamefont
  {Wilczek}},\ }\href {\doibase 10.1103/PhysRevLett.40.279} {\bibfield
  {journal} {\bibinfo  {journal} {Phys. Rev. Lett.}\ }\textbf {\bibinfo
  {volume} {40}},\ \bibinfo {pages} {279} (\bibinfo {year} {1978})}\BibitemShut
  {NoStop}%
\bibitem [{\citenamefont {Peccei}\ and\ \citenamefont
  {Quinn}(1977)}]{Peccei_Quinn77}%
  \BibitemOpen
  \bibfield  {author} {\bibinfo {author} {\bibfnamefont {R.~D.}\ \bibnamefont
  {Peccei}}\ and\ \bibinfo {author} {\bibfnamefont {H.~R.}\ \bibnamefont
  {Quinn}},\ }\href {\doibase 10.1103/PhysRevLett.38.1440} {\bibfield
  {journal} {\bibinfo  {journal} {Phys. Rev. Lett.}\ }\textbf {\bibinfo
  {volume} {38}},\ \bibinfo {pages} {1440} (\bibinfo {year}
  {1977})}\BibitemShut {NoStop}%
\bibitem [{\citenamefont {Sikivie}(2021)}]{Sikivie21}%
  \BibitemOpen
  \bibfield  {author} {\bibinfo {author} {\bibfnamefont {P.}~\bibnamefont
  {Sikivie}},\ }\href {\doibase 10.1103/revmodphys.93.015004} {\bibfield
  {journal} {\bibinfo  {journal} {Reviews of Modern Physics}\ }\textbf
  {\bibinfo {volume} {93}} (\bibinfo {year} {2021}),\
  10.1103/revmodphys.93.015004}\BibitemShut {NoStop}%
\bibitem [{\citenamefont {Arias}\ \emph {et~al.}(2012)\citenamefont {Arias},
  \citenamefont {Cadamuro}, \citenamefont {Goodsell}, \citenamefont {Jaeckel},
  \citenamefont {Redondo},\ and\ \citenamefont {Ringwald}}]{Arias}%
  \BibitemOpen
  \bibfield  {author} {\bibinfo {author} {\bibfnamefont {P.}~\bibnamefont
  {Arias}}, \bibinfo {author} {\bibfnamefont {D.}~\bibnamefont {Cadamuro}},
  \bibinfo {author} {\bibfnamefont {M.}~\bibnamefont {Goodsell}}, \bibinfo
  {author} {\bibfnamefont {J.}~\bibnamefont {Jaeckel}}, \bibinfo {author}
  {\bibfnamefont {J.}~\bibnamefont {Redondo}}, \ and\ \bibinfo {author}
  {\bibfnamefont {A.}~\bibnamefont {Ringwald}},\ }\href {\doibase
  10.1088/1475-7516/2012/06/013} {\bibfield  {journal} {\bibinfo  {journal}
  {Journal of Cosmology and Astroparticle Physics}\ }\textbf {\bibinfo {volume}
  {2012}},\ \bibinfo {pages} {013} (\bibinfo {year} {2012})}\BibitemShut
  {NoStop}%
\bibitem [{\citenamefont {Svrcek}\ and\ \citenamefont
  {Witten}(2006)}]{Svrcek06}%
  \BibitemOpen
  \bibfield  {author} {\bibinfo {author} {\bibfnamefont {P.}~\bibnamefont
  {Svrcek}}\ and\ \bibinfo {author} {\bibfnamefont {E.}~\bibnamefont
  {Witten}},\ }\href {\doibase 10.1088/1126-6708/2006/06/051} {\bibfield
  {journal} {\bibinfo  {journal} {Journal of High Energy Physics}\ }\textbf
  {\bibinfo {volume} {2006}},\ \bibinfo {pages} {051} (\bibinfo {year}
  {2006})}\BibitemShut {NoStop}%
\bibitem [{\citenamefont {Obata}\ \emph {et~al.}(2018)\citenamefont {Obata},
  \citenamefont {Fujita},\ and\ \citenamefont {Michimura}}]{Obata18}%
  \BibitemOpen
  \bibfield  {author} {\bibinfo {author} {\bibfnamefont {I.}~\bibnamefont
  {Obata}}, \bibinfo {author} {\bibfnamefont {T.}~\bibnamefont {Fujita}}, \
  and\ \bibinfo {author} {\bibfnamefont {Y.}~\bibnamefont {Michimura}},\ }\href
  {\doibase 10.1103/PhysRevLett.121.161301} {\bibfield  {journal} {\bibinfo
  {journal} {Phys. Rev. Lett.}\ }\textbf {\bibinfo {volume} {121}},\ \bibinfo
  {pages} {161301} (\bibinfo {year} {2018})}\BibitemShut {NoStop}%
\bibitem [{\citenamefont {Michimura}\ \emph {et~al.}(2020)\citenamefont
  {Michimura}, \citenamefont {Oshima}, \citenamefont {Watanabe}, \citenamefont
  {Kawasaki}, \citenamefont {Takeda}, \citenamefont {Ando}, \citenamefont
  {Nagano}, \citenamefont {Obata},\ and\ \citenamefont {Fujita}}]{Michimura20}%
  \BibitemOpen
  \bibfield  {author} {\bibinfo {author} {\bibfnamefont {Y.}~\bibnamefont
  {Michimura}}, \bibinfo {author} {\bibfnamefont {Y.}~\bibnamefont {Oshima}},
  \bibinfo {author} {\bibfnamefont {T.}~\bibnamefont {Watanabe}}, \bibinfo
  {author} {\bibfnamefont {T.}~\bibnamefont {Kawasaki}}, \bibinfo {author}
  {\bibfnamefont {H.}~\bibnamefont {Takeda}}, \bibinfo {author} {\bibfnamefont
  {M.}~\bibnamefont {Ando}}, \bibinfo {author} {\bibfnamefont {K.}~\bibnamefont
  {Nagano}}, \bibinfo {author} {\bibfnamefont {I.}~\bibnamefont {Obata}}, \
  and\ \bibinfo {author} {\bibfnamefont {T.}~\bibnamefont {Fujita}},\ }\href
  {\doibase 10.1088/1742-6596/1468/1/012032} {\bibfield  {journal} {\bibinfo
  {journal} {Journal of Physics: Conference Series}\ }\textbf {\bibinfo
  {volume} {1468}},\ \bibinfo {pages} {012032} (\bibinfo {year}
  {2020})}\BibitemShut {NoStop}%
\bibitem [{\citenamefont {Nagano}\ \emph {et~al.}(2019)\citenamefont {Nagano},
  \citenamefont {Fujita}, \citenamefont {Michimura},\ and\ \citenamefont
  {Obata}}]{Nagano19}%
  \BibitemOpen
  \bibfield  {author} {\bibinfo {author} {\bibfnamefont {K.}~\bibnamefont
  {Nagano}}, \bibinfo {author} {\bibfnamefont {T.}~\bibnamefont {Fujita}},
  \bibinfo {author} {\bibfnamefont {Y.}~\bibnamefont {Michimura}}, \ and\
  \bibinfo {author} {\bibfnamefont {I.}~\bibnamefont {Obata}},\ }\href
  {\doibase 10.1103/PhysRevLett.123.111301} {\bibfield  {journal} {\bibinfo
  {journal} {Phys. Rev. Lett.}\ }\textbf {\bibinfo {volume} {123}},\ \bibinfo
  {pages} {111301} (\bibinfo {year} {2019})}\BibitemShut {NoStop}%
\bibitem [{\citenamefont {Colpi}\ \emph {et~al.}(2024)\citenamefont {Colpi},
  \citenamefont {Danzmann}, \citenamefont {Hewitson}, \citenamefont
  {Holley-Bockelmann}, \citenamefont {Jetzer}, \citenamefont {Nelemans},
  \citenamefont {Petiteau}, \citenamefont {Shoemaker}, \citenamefont
  {Sopuerta}, \citenamefont {Stebbins}, \citenamefont {Tanvir}, \citenamefont
  {Ward}, \citenamefont {Weber}, \citenamefont {Thorpe}, \citenamefont
  {Daurskikh}, \citenamefont {Deep}, \citenamefont {Núñez}, \citenamefont
  {Marirrodriga}, \citenamefont {Gehler}, \citenamefont {Halain}, \citenamefont
  {Jennrich}, \citenamefont {Lammers}, \citenamefont {Larrañaga},
  \citenamefont {Lieser}, \citenamefont {Lützgendorf}, \citenamefont
  {Martens}, \citenamefont {Mondin}, \citenamefont {Niño}, \citenamefont
  {Amaro-Seoane}, \citenamefont {Sedda}, \citenamefont {Auclair}, \citenamefont
  {Babak}, \citenamefont {Baghi}, \citenamefont {Baibhav}, \citenamefont
  {Baker}, \citenamefont {Bayle}, \citenamefont {Berry}, \citenamefont {Berti},
  \citenamefont {Boileau}, \citenamefont {Bonetti}, \citenamefont {Brito},
  \citenamefont {Buscicchio}, \citenamefont {Calcagni}, \citenamefont {Capelo},
  \citenamefont {Caprini}, \citenamefont {Caputo}, \citenamefont {Castelli},
  \citenamefont {Chen}, \citenamefont {Chen}, \citenamefont {Chua},
  \citenamefont {Davies}, \citenamefont {Derdzinski}, \citenamefont {Domcke},
  \citenamefont {Doneva}, \citenamefont {Dvorkin}, \citenamefont {Ezquiaga},
  \citenamefont {Gair}, \citenamefont {Haiman}, \citenamefont {Harry},
  \citenamefont {Hartwig}, \citenamefont {Hees}, \citenamefont {Heffernan},
  \citenamefont {Husa}, \citenamefont {Izquierdo}, \citenamefont {Karnesis},
  \citenamefont {Klein}, \citenamefont {Korol}, \citenamefont {Korsakova},
  \citenamefont {Kupfer}, \citenamefont {Laghi}, \citenamefont {Lamberts},
  \citenamefont {Larson}, \citenamefont {Jeune}, \citenamefont {Lewicki},
  \citenamefont {Littenberg}, \citenamefont {Madge}, \citenamefont {Mangiagli},
  \citenamefont {Marsat}, \citenamefont {Vilchez}, \citenamefont {Maselli},
  \citenamefont {Mathews}, \citenamefont {van~de Meent}, \citenamefont
  {Muratore}, \citenamefont {Nardini}, \citenamefont {Pani}, \citenamefont
  {Peloso}, \citenamefont {Pieroni}, \citenamefont {Pound}, \citenamefont
  {Quelquejay-Leclere}, \citenamefont {Ricciardone}, \citenamefont {Rossi},
  \citenamefont {Sartirana}, \citenamefont {Savalle}, \citenamefont {Sberna},
  \citenamefont {Sesana}, \citenamefont {Shoemaker}, \citenamefont {Slutsky},
  \citenamefont {Sotiriou}, \citenamefont {Speri}, \citenamefont {Staab},
  \citenamefont {Steer}, \citenamefont {Tamanini}, \citenamefont {Tasinato},
  \citenamefont {Torrado}, \citenamefont {Torres-Orjuela}, \citenamefont
  {Toubiana}, \citenamefont {Vallisneri}, \citenamefont {Vecchio},
  \citenamefont {Volonteri}, \citenamefont {Yagi},\ and\ \citenamefont
  {Zwick}}]{Colpi24}%
  \BibitemOpen
  \bibfield  {author} {\bibinfo {author} {\bibfnamefont {M.}~\bibnamefont
  {Colpi}}, \bibinfo {author} {\bibfnamefont {K.}~\bibnamefont {Danzmann}},
  \bibinfo {author} {\bibfnamefont {M.}~\bibnamefont {Hewitson}}, \bibinfo
  {author} {\bibfnamefont {K.}~\bibnamefont {Holley-Bockelmann}}, \bibinfo
  {author} {\bibfnamefont {P.}~\bibnamefont {Jetzer}}, \bibinfo {author}
  {\bibfnamefont {G.}~\bibnamefont {Nelemans}}, \bibinfo {author}
  {\bibfnamefont {A.}~\bibnamefont {Petiteau}}, \bibinfo {author}
  {\bibfnamefont {D.}~\bibnamefont {Shoemaker}}, \bibinfo {author}
  {\bibfnamefont {C.}~\bibnamefont {Sopuerta}}, \bibinfo {author}
  {\bibfnamefont {R.}~\bibnamefont {Stebbins}}, \bibinfo {author}
  {\bibfnamefont {N.}~\bibnamefont {Tanvir}}, \bibinfo {author} {\bibfnamefont
  {H.}~\bibnamefont {Ward}}, \bibinfo {author} {\bibfnamefont {W.~J.}\
  \bibnamefont {Weber}}, \bibinfo {author} {\bibfnamefont {I.}~\bibnamefont
  {Thorpe}}, \bibinfo {author} {\bibfnamefont {A.}~\bibnamefont {Daurskikh}},
  \bibinfo {author} {\bibfnamefont {A.}~\bibnamefont {Deep}}, \bibinfo {author}
  {\bibfnamefont {I.~F.}\ \bibnamefont {Núñez}}, \bibinfo {author}
  {\bibfnamefont {C.~G.}\ \bibnamefont {Marirrodriga}}, \bibinfo {author}
  {\bibfnamefont {M.}~\bibnamefont {Gehler}}, \bibinfo {author} {\bibfnamefont
  {J.-P.}\ \bibnamefont {Halain}}, \bibinfo {author} {\bibfnamefont
  {O.}~\bibnamefont {Jennrich}}, \bibinfo {author} {\bibfnamefont
  {U.}~\bibnamefont {Lammers}}, \bibinfo {author} {\bibfnamefont
  {J.}~\bibnamefont {Larrañaga}}, \bibinfo {author} {\bibfnamefont
  {M.}~\bibnamefont {Lieser}}, \bibinfo {author} {\bibfnamefont
  {N.}~\bibnamefont {Lützgendorf}}, \bibinfo {author} {\bibfnamefont
  {W.}~\bibnamefont {Martens}}, \bibinfo {author} {\bibfnamefont
  {L.}~\bibnamefont {Mondin}}, \bibinfo {author} {\bibfnamefont {A.~P.}\
  \bibnamefont {Niño}}, \bibinfo {author} {\bibfnamefont {P.}~\bibnamefont
  {Amaro-Seoane}}, \bibinfo {author} {\bibfnamefont {M.~A.}\ \bibnamefont
  {Sedda}}, \bibinfo {author} {\bibfnamefont {P.}~\bibnamefont {Auclair}},
  \bibinfo {author} {\bibfnamefont {S.}~\bibnamefont {Babak}}, \bibinfo
  {author} {\bibfnamefont {Q.}~\bibnamefont {Baghi}}, \bibinfo {author}
  {\bibfnamefont {V.}~\bibnamefont {Baibhav}}, \bibinfo {author} {\bibfnamefont
  {T.}~\bibnamefont {Baker}}, \bibinfo {author} {\bibfnamefont {J.-B.}\
  \bibnamefont {Bayle}}, \bibinfo {author} {\bibfnamefont {C.}~\bibnamefont
  {Berry}}, \bibinfo {author} {\bibfnamefont {E.}~\bibnamefont {Berti}},
  \bibinfo {author} {\bibfnamefont {G.}~\bibnamefont {Boileau}}, \bibinfo
  {author} {\bibfnamefont {M.}~\bibnamefont {Bonetti}}, \bibinfo {author}
  {\bibfnamefont {R.}~\bibnamefont {Brito}}, \bibinfo {author} {\bibfnamefont
  {R.}~\bibnamefont {Buscicchio}}, \bibinfo {author} {\bibfnamefont
  {G.}~\bibnamefont {Calcagni}}, \bibinfo {author} {\bibfnamefont {P.~R.}\
  \bibnamefont {Capelo}}, \bibinfo {author} {\bibfnamefont {C.}~\bibnamefont
  {Caprini}}, \bibinfo {author} {\bibfnamefont {A.}~\bibnamefont {Caputo}},
  \bibinfo {author} {\bibfnamefont {E.}~\bibnamefont {Castelli}}, \bibinfo
  {author} {\bibfnamefont {H.-Y.}\ \bibnamefont {Chen}}, \bibinfo {author}
  {\bibfnamefont {X.}~\bibnamefont {Chen}}, \bibinfo {author} {\bibfnamefont
  {A.}~\bibnamefont {Chua}}, \bibinfo {author} {\bibfnamefont {G.}~\bibnamefont
  {Davies}}, \bibinfo {author} {\bibfnamefont {A.}~\bibnamefont {Derdzinski}},
  \bibinfo {author} {\bibfnamefont {V.~F.}\ \bibnamefont {Domcke}}, \bibinfo
  {author} {\bibfnamefont {D.}~\bibnamefont {Doneva}}, \bibinfo {author}
  {\bibfnamefont {I.}~\bibnamefont {Dvorkin}}, \bibinfo {author} {\bibfnamefont
  {J.~M.}\ \bibnamefont {Ezquiaga}}, \bibinfo {author} {\bibfnamefont
  {J.}~\bibnamefont {Gair}}, \bibinfo {author} {\bibfnamefont {Z.}~\bibnamefont
  {Haiman}}, \bibinfo {author} {\bibfnamefont {I.}~\bibnamefont {Harry}},
  \bibinfo {author} {\bibfnamefont {O.}~\bibnamefont {Hartwig}}, \bibinfo
  {author} {\bibfnamefont {A.}~\bibnamefont {Hees}}, \bibinfo {author}
  {\bibfnamefont {A.}~\bibnamefont {Heffernan}}, \bibinfo {author}
  {\bibfnamefont {S.}~\bibnamefont {Husa}}, \bibinfo {author} {\bibfnamefont
  {D.}~\bibnamefont {Izquierdo}}, \bibinfo {author} {\bibfnamefont
  {N.}~\bibnamefont {Karnesis}}, \bibinfo {author} {\bibfnamefont
  {A.}~\bibnamefont {Klein}}, \bibinfo {author} {\bibfnamefont
  {V.}~\bibnamefont {Korol}}, \bibinfo {author} {\bibfnamefont
  {N.}~\bibnamefont {Korsakova}}, \bibinfo {author} {\bibfnamefont
  {T.}~\bibnamefont {Kupfer}}, \bibinfo {author} {\bibfnamefont
  {D.}~\bibnamefont {Laghi}}, \bibinfo {author} {\bibfnamefont
  {A.}~\bibnamefont {Lamberts}}, \bibinfo {author} {\bibfnamefont
  {S.}~\bibnamefont {Larson}}, \bibinfo {author} {\bibfnamefont {M.~L.}\
  \bibnamefont {Jeune}}, \bibinfo {author} {\bibfnamefont {M.}~\bibnamefont
  {Lewicki}}, \bibinfo {author} {\bibfnamefont {T.}~\bibnamefont {Littenberg}},
  \bibinfo {author} {\bibfnamefont {E.}~\bibnamefont {Madge}}, \bibinfo
  {author} {\bibfnamefont {A.}~\bibnamefont {Mangiagli}}, \bibinfo {author}
  {\bibfnamefont {S.}~\bibnamefont {Marsat}}, \bibinfo {author} {\bibfnamefont
  {I.~M.}\ \bibnamefont {Vilchez}}, \bibinfo {author} {\bibfnamefont
  {A.}~\bibnamefont {Maselli}}, \bibinfo {author} {\bibfnamefont
  {J.}~\bibnamefont {Mathews}}, \bibinfo {author} {\bibfnamefont
  {M.}~\bibnamefont {van~de Meent}}, \bibinfo {author} {\bibfnamefont
  {M.}~\bibnamefont {Muratore}}, \bibinfo {author} {\bibfnamefont
  {G.}~\bibnamefont {Nardini}}, \bibinfo {author} {\bibfnamefont
  {P.}~\bibnamefont {Pani}}, \bibinfo {author} {\bibfnamefont {M.}~\bibnamefont
  {Peloso}}, \bibinfo {author} {\bibfnamefont {M.}~\bibnamefont {Pieroni}},
  \bibinfo {author} {\bibfnamefont {A.}~\bibnamefont {Pound}}, \bibinfo
  {author} {\bibfnamefont {H.}~\bibnamefont {Quelquejay-Leclere}}, \bibinfo
  {author} {\bibfnamefont {A.}~\bibnamefont {Ricciardone}}, \bibinfo {author}
  {\bibfnamefont {E.~M.}\ \bibnamefont {Rossi}}, \bibinfo {author}
  {\bibfnamefont {A.}~\bibnamefont {Sartirana}}, \bibinfo {author}
  {\bibfnamefont {E.}~\bibnamefont {Savalle}}, \bibinfo {author} {\bibfnamefont
  {L.}~\bibnamefont {Sberna}}, \bibinfo {author} {\bibfnamefont
  {A.}~\bibnamefont {Sesana}}, \bibinfo {author} {\bibfnamefont
  {D.}~\bibnamefont {Shoemaker}}, \bibinfo {author} {\bibfnamefont
  {J.}~\bibnamefont {Slutsky}}, \bibinfo {author} {\bibfnamefont
  {T.}~\bibnamefont {Sotiriou}}, \bibinfo {author} {\bibfnamefont
  {L.}~\bibnamefont {Speri}}, \bibinfo {author} {\bibfnamefont
  {M.}~\bibnamefont {Staab}}, \bibinfo {author} {\bibfnamefont
  {D.}~\bibnamefont {Steer}}, \bibinfo {author} {\bibfnamefont
  {N.}~\bibnamefont {Tamanini}}, \bibinfo {author} {\bibfnamefont
  {G.}~\bibnamefont {Tasinato}}, \bibinfo {author} {\bibfnamefont
  {J.}~\bibnamefont {Torrado}}, \bibinfo {author} {\bibfnamefont
  {A.}~\bibnamefont {Torres-Orjuela}}, \bibinfo {author} {\bibfnamefont
  {A.}~\bibnamefont {Toubiana}}, \bibinfo {author} {\bibfnamefont
  {M.}~\bibnamefont {Vallisneri}}, \bibinfo {author} {\bibfnamefont
  {A.}~\bibnamefont {Vecchio}}, \bibinfo {author} {\bibfnamefont
  {M.}~\bibnamefont {Volonteri}}, \bibinfo {author} {\bibfnamefont
  {K.}~\bibnamefont {Yagi}}, \ and\ \bibinfo {author} {\bibfnamefont
  {L.}~\bibnamefont {Zwick}},\ }\href@noop {} {\enquote {\bibinfo {title} {Lisa
  definition study report},}\ } (\bibinfo {year} {2024}),\ \Eprint
  {http://arxiv.org/abs/2402.07571} {arXiv:2402.07571 [astro-ph.CO]}
  \BibitemShut {NoStop}%
\bibitem [{\citenamefont {Luo}\ \emph {et~al.}(2016)\citenamefont {Luo},
  \citenamefont {Chen}, \citenamefont {Duan}, \citenamefont {Gong},
  \citenamefont {Hu}, \citenamefont {Ji}, \citenamefont {Liu}, \citenamefont
  {Mei}, \citenamefont {Milyukov}, \citenamefont {Sazhin}, \citenamefont
  {Shao}, \citenamefont {Toth}, \citenamefont {Tu}, \citenamefont {Wang},
  \citenamefont {Wang}, \citenamefont {Yeh}, \citenamefont {Zhan},
  \citenamefont {Zhang}, \citenamefont {Zharov},\ and\ \citenamefont
  {Zhou}}]{Luo16}%
  \BibitemOpen
  \bibfield  {author} {\bibinfo {author} {\bibfnamefont {J.}~\bibnamefont
  {Luo}}, \bibinfo {author} {\bibfnamefont {L.-S.}\ \bibnamefont {Chen}},
  \bibinfo {author} {\bibfnamefont {H.-Z.}\ \bibnamefont {Duan}}, \bibinfo
  {author} {\bibfnamefont {Y.-G.}\ \bibnamefont {Gong}}, \bibinfo {author}
  {\bibfnamefont {S.}~\bibnamefont {Hu}}, \bibinfo {author} {\bibfnamefont
  {J.}~\bibnamefont {Ji}}, \bibinfo {author} {\bibfnamefont {Q.}~\bibnamefont
  {Liu}}, \bibinfo {author} {\bibfnamefont {J.}~\bibnamefont {Mei}}, \bibinfo
  {author} {\bibfnamefont {V.}~\bibnamefont {Milyukov}}, \bibinfo {author}
  {\bibfnamefont {M.}~\bibnamefont {Sazhin}}, \bibinfo {author} {\bibfnamefont
  {C.-G.}\ \bibnamefont {Shao}}, \bibinfo {author} {\bibfnamefont {V.~T.}\
  \bibnamefont {Toth}}, \bibinfo {author} {\bibfnamefont {H.-B.}\ \bibnamefont
  {Tu}}, \bibinfo {author} {\bibfnamefont {Y.}~\bibnamefont {Wang}}, \bibinfo
  {author} {\bibfnamefont {Y.}~\bibnamefont {Wang}}, \bibinfo {author}
  {\bibfnamefont {H.-C.}\ \bibnamefont {Yeh}}, \bibinfo {author} {\bibfnamefont
  {M.-S.}\ \bibnamefont {Zhan}}, \bibinfo {author} {\bibfnamefont
  {Y.}~\bibnamefont {Zhang}}, \bibinfo {author} {\bibfnamefont
  {V.}~\bibnamefont {Zharov}}, \ and\ \bibinfo {author} {\bibfnamefont {Z.-B.}\
  \bibnamefont {Zhou}},\ }\href {\doibase 10.1088/0264-9381/33/3/035010}
  {\bibfield  {journal} {\bibinfo  {journal} {Classical and Quantum Gravity}\
  }\textbf {\bibinfo {volume} {33}},\ \bibinfo {pages} {035010} (\bibinfo
  {year} {2016})}\BibitemShut {NoStop}%
\bibitem [{\citenamefont {Luo}\ \emph {et~al.}(2020)\citenamefont {Luo},
  \citenamefont {Guo}, \citenamefont {Jin}, \citenamefont {Wu},\ and\
  \citenamefont {Hu}}]{Luo20}%
  \BibitemOpen
  \bibfield  {author} {\bibinfo {author} {\bibfnamefont {Z.}~\bibnamefont
  {Luo}}, \bibinfo {author} {\bibfnamefont {Z.}~\bibnamefont {Guo}}, \bibinfo
  {author} {\bibfnamefont {G.}~\bibnamefont {Jin}}, \bibinfo {author}
  {\bibfnamefont {Y.}~\bibnamefont {Wu}}, \ and\ \bibinfo {author}
  {\bibfnamefont {W.}~\bibnamefont {Hu}},\ }\href {\doibase
  https://doi.org/10.1016/j.rinp.2019.102918} {\bibfield  {journal} {\bibinfo
  {journal} {Results in Physics}\ }\textbf {\bibinfo {volume} {16}},\ \bibinfo
  {pages} {102918} (\bibinfo {year} {2020})}\BibitemShut {NoStop}%
\bibitem [{\citenamefont {Crowder}\ and\ \citenamefont
  {Cornish}(2005)}]{Crowder05}%
  \BibitemOpen
  \bibfield  {author} {\bibinfo {author} {\bibfnamefont {J.}~\bibnamefont
  {Crowder}}\ and\ \bibinfo {author} {\bibfnamefont {N.~J.}\ \bibnamefont
  {Cornish}},\ }\href {\doibase 10.1103/PhysRevD.72.083005} {\bibfield
  {journal} {\bibinfo  {journal} {Phys. Rev. D}\ }\textbf {\bibinfo {volume}
  {72}},\ \bibinfo {pages} {083005} (\bibinfo {year} {2005})}\BibitemShut
  {NoStop}%
\bibitem [{\citenamefont {Armstrong}\ \emph {et~al.}(1999)\citenamefont
  {Armstrong}, \citenamefont {Estabrook},\ and\ \citenamefont
  {Tinto}}]{Armstrong99}%
  \BibitemOpen
  \bibfield  {author} {\bibinfo {author} {\bibfnamefont {J.~W.}\ \bibnamefont
  {Armstrong}}, \bibinfo {author} {\bibfnamefont {F.~B.}\ \bibnamefont
  {Estabrook}}, \ and\ \bibinfo {author} {\bibfnamefont {M.}~\bibnamefont
  {Tinto}},\ }\href {\doibase 10.1086/308110} {\bibfield  {journal} {\bibinfo
  {journal} {The Astrophysical Journal}\ }\textbf {\bibinfo {volume} {527}},\
  \bibinfo {pages} {814} (\bibinfo {year} {1999})}\BibitemShut {NoStop}%
\bibitem [{\citenamefont {Tinto}\ and\ \citenamefont
  {Armstrong}(1999)}]{Tinto99}%
  \BibitemOpen
  \bibfield  {author} {\bibinfo {author} {\bibfnamefont {M.}~\bibnamefont
  {Tinto}}\ and\ \bibinfo {author} {\bibfnamefont {J.~W.}\ \bibnamefont
  {Armstrong}},\ }\href {\doibase 10.1103/PhysRevD.59.102003} {\bibfield
  {journal} {\bibinfo  {journal} {Phys. Rev. D}\ }\textbf {\bibinfo {volume}
  {59}},\ \bibinfo {pages} {102003} (\bibinfo {year} {1999})}\BibitemShut
  {NoStop}%
\bibitem [{\citenamefont {Dhurandhar}\ \emph {et~al.}(2002)\citenamefont
  {Dhurandhar}, \citenamefont {Nayak},\ and\ \citenamefont
  {Vinet}}]{Dhurandhar02}%
  \BibitemOpen
  \bibfield  {author} {\bibinfo {author} {\bibfnamefont {S.~V.}\ \bibnamefont
  {Dhurandhar}}, \bibinfo {author} {\bibfnamefont {K.~R.}\ \bibnamefont
  {Nayak}}, \ and\ \bibinfo {author} {\bibfnamefont {J.-Y.}\ \bibnamefont
  {Vinet}},\ }\href {\doibase 10.1103/physrevd.65.102002} {\bibfield  {journal}
  {\bibinfo  {journal} {Physical Review D}\ }\textbf {\bibinfo {volume} {65}}
  (\bibinfo {year} {2002}),\ 10.1103/physrevd.65.102002}\BibitemShut {NoStop}%
\bibitem [{\citenamefont {{Foster}}\ \emph {et~al.}(2018)\citenamefont
  {{Foster}}, \citenamefont {{Rodd}},\ and\ \citenamefont
  {{Safdi}}}]{foster:2018aa}%
  \BibitemOpen
  \bibfield  {author} {\bibinfo {author} {\bibfnamefont {J.~W.}\ \bibnamefont
  {{Foster}}}, \bibinfo {author} {\bibfnamefont {N.~L.}\ \bibnamefont
  {{Rodd}}}, \ and\ \bibinfo {author} {\bibfnamefont {B.~R.}\ \bibnamefont
  {{Safdi}}},\ }\href {\doibase 10.1103/PhysRevD.97.123006} {\bibfield
  {journal} {\bibinfo  {journal} {Physical Review D}\ }\textbf {\bibinfo
  {volume} {97}},\ \bibinfo {eid} {123006} (\bibinfo {year} {2018})},\ \Eprint
  {http://arxiv.org/abs/1711.10489} {arXiv:1711.10489} \BibitemShut {NoStop}%
\bibitem [{\citenamefont {McMillan}(2011)}]{McMillan11}%
  \BibitemOpen
  \bibfield  {author} {\bibinfo {author} {\bibfnamefont {P.~J.}\ \bibnamefont
  {McMillan}},\ }\href {\doibase 10.1111/j.1365-2966.2011.18564.x} {\bibfield
  {journal} {\bibinfo  {journal} {Monthly Notices of the Royal Astronomical
  Society}\ }\textbf {\bibinfo {volume} {414}},\ \bibinfo {pages} {2446}
  (\bibinfo {year} {2011})}\BibitemShut {NoStop}%
\bibitem [{\citenamefont {Cirelli}\ \emph {et~al.}(2024)\citenamefont
  {Cirelli}, \citenamefont {Strumia},\ and\ \citenamefont {Zupan}}]{Cirelli24}%
  \BibitemOpen
  \bibfield  {author} {\bibinfo {author} {\bibfnamefont {M.}~\bibnamefont
  {Cirelli}}, \bibinfo {author} {\bibfnamefont {A.}~\bibnamefont {Strumia}}, \
  and\ \bibinfo {author} {\bibfnamefont {J.}~\bibnamefont {Zupan}},\
  }\href@noop {} {\enquote {\bibinfo {title} {Dark matter},}\ } (\bibinfo
  {year} {2024}),\ \Eprint {http://arxiv.org/abs/2406.01705} {arXiv:2406.01705
  [hep-ph]} \BibitemShut {NoStop}%
\bibitem [{\citenamefont {Evans}\ \emph {et~al.}(2019)\citenamefont {Evans},
  \citenamefont {O'Hare},\ and\ \citenamefont {McCabe}}]{Evans19}%
  \BibitemOpen
  \bibfield  {author} {\bibinfo {author} {\bibfnamefont {N.~W.}\ \bibnamefont
  {Evans}}, \bibinfo {author} {\bibfnamefont {C.~A.~J.}\ \bibnamefont
  {O'Hare}}, \ and\ \bibinfo {author} {\bibfnamefont {C.}~\bibnamefont
  {McCabe}},\ }\href {\doibase 10.1103/PhysRevD.99.023012} {\bibfield
  {journal} {\bibinfo  {journal} {Phys. Rev. D}\ }\textbf {\bibinfo {volume}
  {99}},\ \bibinfo {pages} {023012} (\bibinfo {year} {2019})}\BibitemShut
  {NoStop}%
\bibitem [{\citenamefont {Bayle}\ and\ \citenamefont
  {Hartwig}(2023)}]{Bayle23}%
  \BibitemOpen
  \bibfield  {author} {\bibinfo {author} {\bibfnamefont {J.-B.}\ \bibnamefont
  {Bayle}}\ and\ \bibinfo {author} {\bibfnamefont {O.}~\bibnamefont
  {Hartwig}},\ }\href {\doibase 10.1103/PhysRevD.107.083019} {\bibfield
  {journal} {\bibinfo  {journal} {Phys. Rev. D}\ }\textbf {\bibinfo {volume}
  {107}},\ \bibinfo {pages} {083019} (\bibinfo {year} {2023})}\BibitemShut
  {NoStop}%
\bibitem [{\citenamefont {Heinzel}\ \emph {et~al.}(2024)\citenamefont
  {Heinzel}, \citenamefont {\'Alvarez-Vizoso}, \citenamefont
  {Dovale-\'Alvarez},\ and\ \citenamefont {Wiesner}}]{Heinzel24}%
  \BibitemOpen
  \bibfield  {author} {\bibinfo {author} {\bibfnamefont {G.}~\bibnamefont
  {Heinzel}}, \bibinfo {author} {\bibfnamefont {J.}~\bibnamefont
  {\'Alvarez-Vizoso}}, \bibinfo {author} {\bibfnamefont {M.}~\bibnamefont
  {Dovale-\'Alvarez}}, \ and\ \bibinfo {author} {\bibfnamefont
  {K.}~\bibnamefont {Wiesner}},\ }\href {\doibase 10.1103/PhysRevD.110.042002}
  {\bibfield  {journal} {\bibinfo  {journal} {Phys. Rev. D}\ }\textbf {\bibinfo
  {volume} {110}},\ \bibinfo {pages} {042002} (\bibinfo {year}
  {2024})}\BibitemShut {NoStop}%
\bibitem [{\citenamefont {Cornish}\ and\ \citenamefont
  {Rubbo}(2003)}]{Cornish03}%
  \BibitemOpen
  \bibfield  {author} {\bibinfo {author} {\bibfnamefont {N.~J.}\ \bibnamefont
  {Cornish}}\ and\ \bibinfo {author} {\bibfnamefont {L.~J.}\ \bibnamefont
  {Rubbo}},\ }\href {\doibase 10.1103/PhysRevD.67.022001} {\bibfield  {journal}
  {\bibinfo  {journal} {Phys. Rev. D}\ }\textbf {\bibinfo {volume} {67}},\
  \bibinfo {pages} {022001} (\bibinfo {year} {2003})}\BibitemShut {NoStop}%
\bibitem [{\citenamefont {Yu}\ \emph {et~al.}(2023)\citenamefont {Yu},
  \citenamefont {Yao}, \citenamefont {Tang},\ and\ \citenamefont {Wu}}]{Yu23}%
  \BibitemOpen
  \bibfield  {author} {\bibinfo {author} {\bibfnamefont {J.-C.}\ \bibnamefont
  {Yu}}, \bibinfo {author} {\bibfnamefont {Y.-H.}\ \bibnamefont {Yao}},
  \bibinfo {author} {\bibfnamefont {Y.}~\bibnamefont {Tang}}, \ and\ \bibinfo
  {author} {\bibfnamefont {Y.-L.}\ \bibnamefont {Wu}},\ }\href {\doibase
  10.1103/PhysRevD.108.083007} {\bibfield  {journal} {\bibinfo  {journal}
  {Phys. Rev. D}\ }\textbf {\bibinfo {volume} {108}},\ \bibinfo {pages}
  {083007} (\bibinfo {year} {2023})}\BibitemShut {NoStop}%
\bibitem [{\citenamefont {Morisaki}\ and\ \citenamefont
  {Suyama}(2019)}]{Morisaki19}%
  \BibitemOpen
  \bibfield  {author} {\bibinfo {author} {\bibfnamefont {S.}~\bibnamefont
  {Morisaki}}\ and\ \bibinfo {author} {\bibfnamefont {T.}~\bibnamefont
  {Suyama}},\ }\href {\doibase 10.1103/physrevd.100.123512} {\bibfield
  {journal} {\bibinfo  {journal} {Physical Review D}\ }\textbf {\bibinfo
  {volume} {100}} (\bibinfo {year} {2019}),\
  10.1103/physrevd.100.123512}\BibitemShut {NoStop}%
\bibitem [{\citenamefont {{Tinto}}\ and\ \citenamefont
  {{Dhurandhar}}(2021)}]{tinto:2021aa}%
  \BibitemOpen
  \bibfield  {author} {\bibinfo {author} {\bibfnamefont {M.}~\bibnamefont
  {{Tinto}}}\ and\ \bibinfo {author} {\bibfnamefont {S.~V.}\ \bibnamefont
  {{Dhurandhar}}},\ }\href@noop {} {\bibfield  {journal} {\bibinfo  {journal}
  {Living Reviews in Relativity}\ }\textbf {\bibinfo {volume} {24}},\ \bibinfo
  {pages} {1} (\bibinfo {year} {2021})}\BibitemShut {NoStop}%
\bibitem [{\citenamefont {{Hartwig}}\ and\ \citenamefont
  {{Muratore}}(2022)}]{hartwig:2022aa}%
  \BibitemOpen
  \bibfield  {author} {\bibinfo {author} {\bibfnamefont {O.}~\bibnamefont
  {{Hartwig}}}\ and\ \bibinfo {author} {\bibfnamefont {M.}~\bibnamefont
  {{Muratore}}},\ }\href {\doibase 10.1103/PhysRevD.105.062006} {\bibfield
  {journal} {\bibinfo  {journal} {\prd}\ }\textbf {\bibinfo {volume} {105}},\
  \bibinfo {eid} {062006} (\bibinfo {year} {2022})},\ \Eprint
  {http://arxiv.org/abs/2111.00975} {arXiv:2111.00975 [gr-qc]} \BibitemShut
  {NoStop}%
\bibitem [{\citenamefont {Martens}\ and\ \citenamefont
  {Joffre}(2021)}]{Martens21}%
  \BibitemOpen
  \bibfield  {author} {\bibinfo {author} {\bibfnamefont {W.}~\bibnamefont
  {Martens}}\ and\ \bibinfo {author} {\bibfnamefont {E.}~\bibnamefont
  {Joffre}},\ }\href {\doibase 10.1007/s40295-021-00263-2} {\bibfield
  {journal} {\bibinfo  {journal} {The Journal of the Astronautical Sciences}\
  }\textbf {\bibinfo {volume} {68}},\ \bibinfo {pages} {402–443} (\bibinfo
  {year} {2021})}\BibitemShut {NoStop}%
\bibitem [{\citenamefont {Robson}\ \emph {et~al.}(2019)\citenamefont {Robson},
  \citenamefont {Cornish},\ and\ \citenamefont {Liu}}]{Robson19}%
  \BibitemOpen
  \bibfield  {author} {\bibinfo {author} {\bibfnamefont {T.}~\bibnamefont
  {Robson}}, \bibinfo {author} {\bibfnamefont {N.~J.}\ \bibnamefont {Cornish}},
  \ and\ \bibinfo {author} {\bibfnamefont {C.}~\bibnamefont {Liu}},\ }\href
  {\doibase 10.1088/1361-6382/ab1101} {\bibfield  {journal} {\bibinfo
  {journal} {Classical and Quantum Gravity}\ }\textbf {\bibinfo {volume}
  {36}},\ \bibinfo {pages} {105011} (\bibinfo {year} {2019})}\BibitemShut
  {NoStop}%
\bibitem [{\citenamefont {Babak}\ \emph {et~al.}(2021)\citenamefont {Babak},
  \citenamefont {Hewitson},\ and\ \citenamefont {Petiteau}}]{Babak21}%
  \BibitemOpen
  \bibfield  {author} {\bibinfo {author} {\bibfnamefont {S.}~\bibnamefont
  {Babak}}, \bibinfo {author} {\bibfnamefont {M.}~\bibnamefont {Hewitson}}, \
  and\ \bibinfo {author} {\bibfnamefont {A.}~\bibnamefont {Petiteau}},\
  }\href@noop {} {\enquote {\bibinfo {title} {Lisa sensitivity and snr
  calculations},}\ } (\bibinfo {year} {2021}),\ \Eprint
  {http://arxiv.org/abs/2108.01167} {arXiv:2108.01167 [astro-ph.IM]}
  \BibitemShut {NoStop}%
\bibitem [{\citenamefont {Hu}\ and\ \citenamefont {Wu}(2017)}]{Hu17}%
  \BibitemOpen
  \bibfield  {author} {\bibinfo {author} {\bibfnamefont {W.-R.}\ \bibnamefont
  {Hu}}\ and\ \bibinfo {author} {\bibfnamefont {Y.-L.}\ \bibnamefont {Wu}},\
  }\href {\doibase 10.1093/nsr/nwx116} {\bibfield  {journal} {\bibinfo
  {journal} {National Science Review}\ }\textbf {\bibinfo {volume} {4}},\
  \bibinfo {pages} {685–686} (\bibinfo {year} {2017})}\BibitemShut {NoStop}%
\bibitem [{\citenamefont {Corbin}\ and\ \citenamefont
  {Cornish}(2006)}]{Corbin06}%
  \BibitemOpen
  \bibfield  {author} {\bibinfo {author} {\bibfnamefont {V.}~\bibnamefont
  {Corbin}}\ and\ \bibinfo {author} {\bibfnamefont {N.~J.}\ \bibnamefont
  {Cornish}},\ }\href {\doibase 10.1088/0264-9381/23/7/014} {\bibfield
  {journal} {\bibinfo  {journal} {Classical and Quantum Gravity}\ }\textbf
  {\bibinfo {volume} {23}},\ \bibinfo {pages} {2435–2446} (\bibinfo {year}
  {2006})}\BibitemShut {NoStop}%
\bibitem [{\citenamefont {Hartwig}\ \emph {et~al.}(2023)\citenamefont
  {Hartwig}, \citenamefont {Lilley}, \citenamefont {Muratore},\ and\
  \citenamefont {Pieroni}}]{Hartwig23}%
  \BibitemOpen
  \bibfield  {author} {\bibinfo {author} {\bibfnamefont {O.}~\bibnamefont
  {Hartwig}}, \bibinfo {author} {\bibfnamefont {M.}~\bibnamefont {Lilley}},
  \bibinfo {author} {\bibfnamefont {M.}~\bibnamefont {Muratore}}, \ and\
  \bibinfo {author} {\bibfnamefont {M.}~\bibnamefont {Pieroni}},\ }\href
  {\doibase 10.1103/physrevd.107.123531} {\bibfield  {journal} {\bibinfo
  {journal} {Physical Review D}\ }\textbf {\bibinfo {volume} {107}} (\bibinfo
  {year} {2023}),\ 10.1103/physrevd.107.123531}\BibitemShut {NoStop}%
\bibitem [{\citenamefont {Derevianko}(2018)}]{Derevianko18}%
  \BibitemOpen
  \bibfield  {author} {\bibinfo {author} {\bibfnamefont {A.}~\bibnamefont
  {Derevianko}},\ }\href {\doibase 10.1103/PhysRevA.97.042506} {\bibfield
  {journal} {\bibinfo  {journal} {Phys. Rev. A}\ }\textbf {\bibinfo {volume}
  {97}},\ \bibinfo {pages} {042506} (\bibinfo {year} {2018})}\BibitemShut
  {NoStop}%
\bibitem [{\citenamefont {Centers}\ \emph {et~al.}(2021)\citenamefont
  {Centers}, \citenamefont {Blanchard}, \citenamefont {Conrad}, \citenamefont
  {Figueroa}, \citenamefont {Garcon}, \citenamefont {Gramolin}, \citenamefont
  {Kimball}, \citenamefont {Lawson}, \citenamefont {Pelssers}, \citenamefont
  {Smiga}, \citenamefont {Sushkov}, \citenamefont {Wickenbrock}, \citenamefont
  {Budker},\ and\ \citenamefont {Derevianko}}]{Centers21}%
  \BibitemOpen
  \bibfield  {author} {\bibinfo {author} {\bibfnamefont {G.~P.}\ \bibnamefont
  {Centers}}, \bibinfo {author} {\bibfnamefont {J.~W.}\ \bibnamefont
  {Blanchard}}, \bibinfo {author} {\bibfnamefont {J.}~\bibnamefont {Conrad}},
  \bibinfo {author} {\bibfnamefont {N.~L.}\ \bibnamefont {Figueroa}}, \bibinfo
  {author} {\bibfnamefont {A.}~\bibnamefont {Garcon}}, \bibinfo {author}
  {\bibfnamefont {A.~V.}\ \bibnamefont {Gramolin}}, \bibinfo {author}
  {\bibfnamefont {D.~F.~J.}\ \bibnamefont {Kimball}}, \bibinfo {author}
  {\bibfnamefont {M.}~\bibnamefont {Lawson}}, \bibinfo {author} {\bibfnamefont
  {B.}~\bibnamefont {Pelssers}}, \bibinfo {author} {\bibfnamefont {J.~A.}\
  \bibnamefont {Smiga}}, \bibinfo {author} {\bibfnamefont {A.~O.}\ \bibnamefont
  {Sushkov}}, \bibinfo {author} {\bibfnamefont {A.}~\bibnamefont
  {Wickenbrock}}, \bibinfo {author} {\bibfnamefont {D.}~\bibnamefont {Budker}},
  \ and\ \bibinfo {author} {\bibfnamefont {A.}~\bibnamefont {Derevianko}},\
  }\href {\doibase 10.1038/s41467-021-27632-7} {\bibfield  {journal} {\bibinfo
  {journal} {Nature Communications}\ }\textbf {\bibinfo {volume} {12}}
  (\bibinfo {year} {2021}),\ 10.1038/s41467-021-27632-7}\BibitemShut {NoStop}%
\bibitem [{\citenamefont {Budker}\ \emph {et~al.}(2014)\citenamefont {Budker},
  \citenamefont {Graham}, \citenamefont {Ledbetter}, \citenamefont
  {Rajendran},\ and\ \citenamefont {Sushkov}}]{Budker14}%
  \BibitemOpen
  \bibfield  {author} {\bibinfo {author} {\bibfnamefont {D.}~\bibnamefont
  {Budker}}, \bibinfo {author} {\bibfnamefont {P.~W.}\ \bibnamefont {Graham}},
  \bibinfo {author} {\bibfnamefont {M.}~\bibnamefont {Ledbetter}}, \bibinfo
  {author} {\bibfnamefont {S.}~\bibnamefont {Rajendran}}, \ and\ \bibinfo
  {author} {\bibfnamefont {A.~O.}\ \bibnamefont {Sushkov}},\ }\href {\doibase
  10.1103/PhysRevX.4.021030} {\bibfield  {journal} {\bibinfo  {journal} {Phys.
  Rev. X}\ }\textbf {\bibinfo {volume} {4}},\ \bibinfo {pages} {021030}
  (\bibinfo {year} {2014})}\BibitemShut {NoStop}%
\bibitem [{\citenamefont {Quang~Nam}\ \emph {et~al.}(2023)\citenamefont
  {Quang~Nam}, \citenamefont {Martino}, \citenamefont {Lemi\`ere},
  \citenamefont {Petiteau}, \citenamefont {Bayle}, \citenamefont {Hartwig},\
  and\ \citenamefont {Staab}}]{Nam23}%
  \BibitemOpen
  \bibfield  {author} {\bibinfo {author} {\bibfnamefont {D.}~\bibnamefont
  {Quang~Nam}}, \bibinfo {author} {\bibfnamefont {J.}~\bibnamefont {Martino}},
  \bibinfo {author} {\bibfnamefont {Y.}~\bibnamefont {Lemi\`ere}}, \bibinfo
  {author} {\bibfnamefont {A.}~\bibnamefont {Petiteau}}, \bibinfo {author}
  {\bibfnamefont {J.-B.}\ \bibnamefont {Bayle}}, \bibinfo {author}
  {\bibfnamefont {O.}~\bibnamefont {Hartwig}}, \ and\ \bibinfo {author}
  {\bibfnamefont {M.}~\bibnamefont {Staab}},\ }\href {\doibase
  10.1103/PhysRevD.108.082004} {\bibfield  {journal} {\bibinfo  {journal}
  {Phys. Rev. D}\ }\textbf {\bibinfo {volume} {108}},\ \bibinfo {pages}
  {082004} (\bibinfo {year} {2023})}\BibitemShut {NoStop}%
\bibitem [{\citenamefont {Cutler}\ and\ \citenamefont {Holz}(2009)}]{Cutler09}%
  \BibitemOpen
  \bibfield  {author} {\bibinfo {author} {\bibfnamefont {C.}~\bibnamefont
  {Cutler}}\ and\ \bibinfo {author} {\bibfnamefont {D.~E.}\ \bibnamefont
  {Holz}},\ }\href {\doibase 10.1103/physrevd.80.104009} {\bibfield  {journal}
  {\bibinfo  {journal} {Physical Review D}\ }\textbf {\bibinfo {volume} {80}}
  (\bibinfo {year} {2009}),\ 10.1103/physrevd.80.104009}\BibitemShut {NoStop}%
\bibitem [{\citenamefont {Anastassopoulos}\ \emph {et~al.}(2017)\citenamefont
  {Anastassopoulos}, \citenamefont {Aune}, \citenamefont {Barth}, \citenamefont
  {Belov}, \citenamefont {Bräuninger}, \citenamefont {Cantatore},
  \citenamefont {Carmona}, \citenamefont {Castel}, \citenamefont {Cetin},
  \citenamefont {Christensen}, \citenamefont {Collar}, \citenamefont {Dafni},
  \citenamefont {Davenport}, \citenamefont {Decker}, \citenamefont {Dermenev},
  \citenamefont {Desch}, \citenamefont {Eleftheriadis}, \citenamefont
  {Fanourakis}, \citenamefont {Ferrer-Ribas}, \citenamefont {Fischer},
  \citenamefont {García}, \citenamefont {Gardikiotis}, \citenamefont {Garza},
  \citenamefont {Gazis}, \citenamefont {Geralis}, \citenamefont {Giomataris},
  \citenamefont {Gninenko}, \citenamefont {Hailey}, \citenamefont {Hasinoff},
  \citenamefont {Hoffmann}, \citenamefont {Iguaz}, \citenamefont {Irastorza},
  \citenamefont {Jakobsen}, \citenamefont {Jacoby}, \citenamefont {Jakovčić},
  \citenamefont {Kaminski}, \citenamefont {Karuza}, \citenamefont {Kralj},
  \citenamefont {Krčmar}, \citenamefont {Kostoglou}, \citenamefont {Krieger},
  \citenamefont {Lakić}, \citenamefont {Laurent}, \citenamefont {Liolios},
  \citenamefont {Ljubičić}, \citenamefont {Luzón}, \citenamefont {Maroudas},
  \citenamefont {Miceli}, \citenamefont {Neff}, \citenamefont {Ortega},
  \citenamefont {Papaevangelou}, \citenamefont {Paraschou}, \citenamefont
  {Pivovaroff}, \citenamefont {Raffelt}, \citenamefont {Rosu}, \citenamefont
  {Ruz}, \citenamefont {Chóliz}, \citenamefont {Savvidis}, \citenamefont
  {Schmidt}, \citenamefont {Semertzidis}, \citenamefont {Solanki},
  \citenamefont {Stewart}, \citenamefont {Vafeiadis}, \citenamefont {Vogel},
  \citenamefont {Yildiz}, \citenamefont {Zioutas},\ and\ \citenamefont
  {Collaboration}}]{CAST}%
  \BibitemOpen
  \bibfield  {author} {\bibinfo {author} {\bibfnamefont {V.}~\bibnamefont
  {Anastassopoulos}}, \bibinfo {author} {\bibfnamefont {S.}~\bibnamefont
  {Aune}}, \bibinfo {author} {\bibfnamefont {K.}~\bibnamefont {Barth}},
  \bibinfo {author} {\bibfnamefont {A.}~\bibnamefont {Belov}}, \bibinfo
  {author} {\bibfnamefont {H.}~\bibnamefont {Bräuninger}}, \bibinfo {author}
  {\bibfnamefont {G.}~\bibnamefont {Cantatore}}, \bibinfo {author}
  {\bibfnamefont {J.~M.}\ \bibnamefont {Carmona}}, \bibinfo {author}
  {\bibfnamefont {J.~F.}\ \bibnamefont {Castel}}, \bibinfo {author}
  {\bibfnamefont {S.~A.}\ \bibnamefont {Cetin}}, \bibinfo {author}
  {\bibfnamefont {F.}~\bibnamefont {Christensen}}, \bibinfo {author}
  {\bibfnamefont {J.~I.}\ \bibnamefont {Collar}}, \bibinfo {author}
  {\bibfnamefont {T.}~\bibnamefont {Dafni}}, \bibinfo {author} {\bibfnamefont
  {M.}~\bibnamefont {Davenport}}, \bibinfo {author} {\bibfnamefont {T.~A.}\
  \bibnamefont {Decker}}, \bibinfo {author} {\bibfnamefont {A.}~\bibnamefont
  {Dermenev}}, \bibinfo {author} {\bibfnamefont {K.}~\bibnamefont {Desch}},
  \bibinfo {author} {\bibfnamefont {C.}~\bibnamefont {Eleftheriadis}}, \bibinfo
  {author} {\bibfnamefont {G.}~\bibnamefont {Fanourakis}}, \bibinfo {author}
  {\bibfnamefont {E.}~\bibnamefont {Ferrer-Ribas}}, \bibinfo {author}
  {\bibfnamefont {H.}~\bibnamefont {Fischer}}, \bibinfo {author} {\bibfnamefont
  {J.~A.}\ \bibnamefont {García}}, \bibinfo {author} {\bibfnamefont
  {A.}~\bibnamefont {Gardikiotis}}, \bibinfo {author} {\bibfnamefont {J.~G.}\
  \bibnamefont {Garza}}, \bibinfo {author} {\bibfnamefont {E.~N.}\ \bibnamefont
  {Gazis}}, \bibinfo {author} {\bibfnamefont {T.}~\bibnamefont {Geralis}},
  \bibinfo {author} {\bibfnamefont {I.}~\bibnamefont {Giomataris}}, \bibinfo
  {author} {\bibfnamefont {S.}~\bibnamefont {Gninenko}}, \bibinfo {author}
  {\bibfnamefont {C.~J.}\ \bibnamefont {Hailey}}, \bibinfo {author}
  {\bibfnamefont {M.~D.}\ \bibnamefont {Hasinoff}}, \bibinfo {author}
  {\bibfnamefont {D.~H.~H.}\ \bibnamefont {Hoffmann}}, \bibinfo {author}
  {\bibfnamefont {F.~J.}\ \bibnamefont {Iguaz}}, \bibinfo {author}
  {\bibfnamefont {I.~G.}\ \bibnamefont {Irastorza}}, \bibinfo {author}
  {\bibfnamefont {A.}~\bibnamefont {Jakobsen}}, \bibinfo {author}
  {\bibfnamefont {J.}~\bibnamefont {Jacoby}}, \bibinfo {author} {\bibfnamefont
  {K.}~\bibnamefont {Jakovčić}}, \bibinfo {author} {\bibfnamefont
  {J.}~\bibnamefont {Kaminski}}, \bibinfo {author} {\bibfnamefont
  {M.}~\bibnamefont {Karuza}}, \bibinfo {author} {\bibfnamefont
  {N.}~\bibnamefont {Kralj}}, \bibinfo {author} {\bibfnamefont
  {M.}~\bibnamefont {Krčmar}}, \bibinfo {author} {\bibfnamefont
  {S.}~\bibnamefont {Kostoglou}}, \bibinfo {author} {\bibfnamefont
  {C.}~\bibnamefont {Krieger}}, \bibinfo {author} {\bibfnamefont
  {B.}~\bibnamefont {Lakić}}, \bibinfo {author} {\bibfnamefont
  {J.}~\bibnamefont {Laurent}}, \bibinfo {author} {\bibfnamefont
  {A.}~\bibnamefont {Liolios}}, \bibinfo {author} {\bibfnamefont
  {A.}~\bibnamefont {Ljubičić}}, \bibinfo {author} {\bibfnamefont
  {G.}~\bibnamefont {Luzón}}, \bibinfo {author} {\bibfnamefont
  {M.}~\bibnamefont {Maroudas}}, \bibinfo {author} {\bibfnamefont
  {L.}~\bibnamefont {Miceli}}, \bibinfo {author} {\bibfnamefont
  {S.}~\bibnamefont {Neff}}, \bibinfo {author} {\bibfnamefont {I.}~\bibnamefont
  {Ortega}}, \bibinfo {author} {\bibfnamefont {T.}~\bibnamefont
  {Papaevangelou}}, \bibinfo {author} {\bibfnamefont {K.}~\bibnamefont
  {Paraschou}}, \bibinfo {author} {\bibfnamefont {M.~J.}\ \bibnamefont
  {Pivovaroff}}, \bibinfo {author} {\bibfnamefont {G.}~\bibnamefont {Raffelt}},
  \bibinfo {author} {\bibfnamefont {M.}~\bibnamefont {Rosu}}, \bibinfo {author}
  {\bibfnamefont {J.}~\bibnamefont {Ruz}}, \bibinfo {author} {\bibfnamefont
  {E.~R.}\ \bibnamefont {Chóliz}}, \bibinfo {author} {\bibfnamefont
  {I.}~\bibnamefont {Savvidis}}, \bibinfo {author} {\bibfnamefont
  {S.}~\bibnamefont {Schmidt}}, \bibinfo {author} {\bibfnamefont
  {Y.}~\bibnamefont {Semertzidis}}, \bibinfo {author} {\bibfnamefont {S.~K.}\
  \bibnamefont {Solanki}}, \bibinfo {author} {\bibfnamefont {L.}~\bibnamefont
  {Stewart}}, \bibinfo {author} {\bibfnamefont {T.}~\bibnamefont {Vafeiadis}},
  \bibinfo {author} {\bibfnamefont {J.~K.}\ \bibnamefont {Vogel}}, \bibinfo
  {author} {\bibfnamefont {S.~C.}\ \bibnamefont {Yildiz}}, \bibinfo {author}
  {\bibfnamefont {K.}~\bibnamefont {Zioutas}}, \ and\ \bibinfo {author}
  {\bibfnamefont {C.}~\bibnamefont {Collaboration}},\ }\href {\doibase
  10.1038/nphys4109} {\bibfield  {journal} {\bibinfo  {journal} {Nature
  Physics}\ }\textbf {\bibinfo {volume} {13}},\ \bibinfo {pages} {584–590}
  (\bibinfo {year} {2017})}\BibitemShut {NoStop}%
\bibitem [{\citenamefont {Payez}\ \emph {et~al.}(2015)\citenamefont {Payez},
  \citenamefont {Evoli}, \citenamefont {Fischer}, \citenamefont {Giannotti},
  \citenamefont {Mirizzi},\ and\ \citenamefont {Ringwald}}]{Payez15}%
  \BibitemOpen
  \bibfield  {author} {\bibinfo {author} {\bibfnamefont {A.}~\bibnamefont
  {Payez}}, \bibinfo {author} {\bibfnamefont {C.}~\bibnamefont {Evoli}},
  \bibinfo {author} {\bibfnamefont {T.}~\bibnamefont {Fischer}}, \bibinfo
  {author} {\bibfnamefont {M.}~\bibnamefont {Giannotti}}, \bibinfo {author}
  {\bibfnamefont {A.}~\bibnamefont {Mirizzi}}, \ and\ \bibinfo {author}
  {\bibfnamefont {A.}~\bibnamefont {Ringwald}},\ }\href {\doibase
  10.1088/1475-7516/2015/02/006} {\bibfield  {journal} {\bibinfo  {journal}
  {Journal of Cosmology and Astroparticle Physics}\ }\textbf {\bibinfo {volume}
  {2015}},\ \bibinfo {pages} {006–006} (\bibinfo {year} {2015})}\BibitemShut
  {NoStop}%
\bibitem [{\citenamefont {O'Hare}(2020)}]{AxionLimits}%
  \BibitemOpen
  \bibfield  {author} {\bibinfo {author} {\bibfnamefont {C.}~\bibnamefont
  {O'Hare}},\ }\href {\doibase 10.5281/zenodo.3932430} {\enquote {\bibinfo
  {title} {cajohare/axionlimits: Axionlimits},}\ }\bibinfo {howpublished}
  {\url{https://cajohare.github.io/AxionLimits/}} (\bibinfo {year}
  {2020})\BibitemShut {NoStop}%
\bibitem [{\citenamefont {Collett}(2005)}]{Collett05}%
  \BibitemOpen
  \bibfield  {author} {\bibinfo {author} {\bibfnamefont {E.}~\bibnamefont
  {Collett}},\ }\href {https://books.google.fr/books?id=5lJwcCsLbLsC} {\emph
  {\bibinfo {title} {Field Guide to Polarization}}},\ Field Guides\ (\bibinfo
  {publisher} {SPIE Press},\ \bibinfo {year} {2005})\BibitemShut {NoStop}%
\bibitem [{\citenamefont {Hecht}(2002)}]{Hecht02}%
  \BibitemOpen
  \bibfield  {author} {\bibinfo {author} {\bibfnamefont {E.}~\bibnamefont
  {Hecht}},\ }\href {https://books.google.fr/books?id=T3ofAQAAMAAJ} {\emph
  {\bibinfo {title} {Optics}}},\ Pearson education\ (\bibinfo  {publisher}
  {Addison-Wesley},\ \bibinfo {year} {2002})\BibitemShut {NoStop}%
\bibitem [{\citenamefont {Fymat}(1971)}]{Fymat71}%
  \BibitemOpen
  \bibfield  {author} {\bibinfo {author} {\bibfnamefont {A.~L.}\ \bibnamefont
  {Fymat}},\ }\href {\doibase 10.1364/AO.10.002711} {\bibfield  {journal}
  {\bibinfo  {journal} {Appl. Opt.}\ }\textbf {\bibinfo {volume} {10}},\
  \bibinfo {pages} {2711} (\bibinfo {year} {1971})}\BibitemShut {NoStop}%
\end{thebibliography}%

\end{document}